\documentclass[preprint,5p,times,twocolumn,authoryear]{elsarticle}

\usepackage{amssymb}
\usepackage{amsmath}
\usepackage{xurl}
\usepackage{tikz}
\usepackage{booktabs} 
\usepackage{tabularx} 
\usepackage{multirow}
\usepackage{tikz} 
\usepackage{fvextra}

\usepackage{tcolorbox}
\tcbuselibrary{breakable}
\newtcolorbox{examplebox}[1]{ colback=blue!5!white, colframe=blue!30!white, fonttitle=\bfseries\small, title={Example: #1}, breakable, left=4pt, right=4pt, top=3pt, bottom=3pt, before upper={\small}, }

\newcommand{\exampleHardCoded}{
\begin{examplebox}{Hand-coded systems}
U.S. state public-benefits eligibility platforms such as Florida's ACCESS, Texas's TIERS, and California's CalWIN are examples of hand-coded systems. Legal scholarship describes these systems as encoding administrative policy into software rules, with caseworkers often reviewing sample outputs before finalisation. Their failures typically arose not from statistical learning but from incorrectly coded rules or policy distortions embedded in the software itself \citep{citronTechnologicalDueProcess2008}.
\end{examplebox}
}

\newcommand{\exampleTransparent}{
\begin{examplebox}{Glass-box systems}
The \emph{Allegheny Family Screening Tool} (AFST), used in Allegheny County, Pennsylvania, scores families' risk of child abuse or neglect on a scale of 1–20 using a logistic regression model trained on historical child welfare records \citep{vaithianathanDevelopingPredictiveModels2017}. Auditors and oversight bodies can inspect the model's coefficients and understand which variables drive higher scores, making the learned logic comparatively legible.
\end{examplebox}
}

\newcommand{\exampleOpaque}{
\begin{examplebox}{Black-box systems}
Amazon's scrapped AI resume screening system serves as an illustrative case for the technical mechanisms of bias in opaque systems. The model was trained on ten years of resumes but developed a pervasive bias against women, penalising applications that alluded to women's colleges or sports \citep{dastinInsightAmazonScraps2018}. Because the system relied on weak, highly contextual proxy signals within an opaque neural architecture, the bias was not located in a single, auditable rule \citep{raghavanMitigatingBiasAlgorithmic2020}. Instead, it emerged from the model's interaction with historical data, demonstrating how opacity shifts the burden of oversight from inspecting legible rules to auditing behavioural outcomes.
\end{examplebox}
}

\newcommand{\exampleFoundation}{
\begin{examplebox}{General-purpose systems}
New York City's ``MyCity'' AI chatbot, launched to help business owners navigate local bureaucracy, illustrates the dangers of externalised evaluation. Powered by external foundation models, the system was deployed to provide legal and regulatory guidance without sufficient domain-specific testing. Consequently, the chatbot frequently hallucinated and advised citizens to break the law---wrongly suggesting, for instance, that employers could legally take a cut of their workers' tips \citep{lecherNYCsAIChatbot2024}. This case highlights the severe risks of assuming a foundation model's general competence will safely transfer to a specific administrative context without rigorous, localised evaluation.
\end{examplebox}
}

\newcommand{\exampleAgentic}{
\begin{examplebox}{Agentic systems}
While agentic systems are still in their infancy, Bürokratt from Estonia offers an early vision. Originally a hand-coded chatbot, Bürokratt is being upgraded with an LLM-based orchestration system to more intelligently handle complex citizen queries by proactively querying information from Estonia's public sector data infrastructure. The long-term vision is for Bürokratt to be an interface to and orchestrator of different government agencies' agent systems \citep{ilvesAgenticStateRethinking2025}.
\end{examplebox}
}

\usepackage[colorlinks=true, linkcolor=blue, citecolor=blue, urlcolor=blue]{hyperref}

\newcommand{\secref}[1]{\S\ref{#1}}
\journal{Government Information Quaterly}

\begin{document}

\begin{frontmatter}




\author[oxford]{Jonathan Rystrøm\fnref{first}\corref{cor1}}
\author[hertie]{Chris Schmitz\fnref{first}}
\author[oxford,harvard]{Nathan Davies}
\author[hertie]{Gerhard Hammerschmid}
\author[utrecht]{Albert Meijer}
\author[oxford]{Chris Russell}

\fntext[first]{These authors contributed equally as joint first authors.}

\cortext[cor1]{Corresponding author. Email: jonathan.rystrom@oii.ox.ac.uk; Tel: +44 01865 287210; Oxford Internet Institute, University of Oxford, Oxford OX2 6GG, United Kingdom.}

\address[oxford]{University of Oxford, Oxford, UK}
\address[hertie]{Hertie School, Berlin, Germany}
\address[utrecht]{Utrecht University, Utrecht, Netherlands}
\address[harvard]{Harvard University, Cambridge, MA, USA}

\title{A Technical Typology of AI Systems in Public Administration}

\begin{abstract}
Research on artificial intelligence (AI) in the public sector often treats ``AI'' as a single category, neglecting technical distinctions between different AI systems. But these distinctions affect how different systems impact core public values like accountability, procedural justice, and non-discrimination. 
This paper argues that public administration research would benefit from more technical precision on ``AI'' and makes three contributions to this end.
First, we introduce a typology of five categories of AI systems: hand-coded, glass-box, black-box, general-purpose, and agentic systems. We calibrate the typology to public administration by grouping system types by their distinct implications for public values. 
Second, we evaluate technical precision in recent public administration research about AI by coding 91 highly-cited papers (2019–2025) using our typology.
We find widespread imprecision: most papers (55\%) leave the studied system underspecified, 31\% motivate their work with a different system than they study, and 41\% make more general conclusions than the studied system supports.
Finally, we give practical recommendations for future research. We highlight common pitfalls to avoid, and suggest that researchers should, at a minimum, provide enough technical detail to locate the studied system in our typology.
To this end, we provide a practical guide -- a short set of diagnostic questions answerable from public information and without specialist technical knowledge.
\end{abstract}





\end{frontmatter}

\section{Introduction} \label{sec:intro}

``AI'' is everywhere in government -- and can refer to almost everything. Consider scholarship on public-sector ``chatbots'' \citep{androutsopoulouTransformingCommunicationCitizens2019}: early chatbots were rule-based systems with fixed, explicitly coded ``conversation trees'' \citep{adamopoulouChatbotsHistoryTechnology2020}, but current systems such as ChatGPT are built on externally developed general-purpose models, can parse ambiguity, and generate fluent, context-sensitive outputs. These systems have different affordances -- such as flexible interaction -- and constraints, such as externalisation of control and unverifiable training data \citep{bommasaniOpportunitiesRisksFoundation2021}. The chatbot is not unusual in this: across benefits eligibility, fraud detection, and case triage, the single label of ``AI'' is routinely used to describe systems with different governance-relevant properties.

For questions of governance, what matters about an AI system is its \emph{affordances} -- what it makes possible or forecloses for a particular actor \citep{zammutoInformationTechnologyChanging2007}.
Even distinguishing `data-driven' AI systems from rule-driven ones, as some research does \citep{wangWhatTypeAlgorithm2023}, can be too imprecise to clarify these affordances. For example, whether an AI system is legible even to experts depends on its technical underpinnings \citep{burrellHowMachineThinks2016}: some remain inspectable while others are black boxes -- a difference in what auditors can examine or citizens can contest. 
Similarly, general-purpose AI systems afford vastly different things than task-specific models, for example by lowering barriers to adoption and externalising control over training data. This shift in affordances was consequential enough that the European Union's AI Act was updated in response \citep{gstreinGeneralpurposeAIRegulation2024, wangDistinguishingTaskspecificGeneralpurpose2026} -- but public administration research has not reflected it clearly.

It is therefore evident that technical precision about ``AI'' can be helpful in public administration research -- but simultaneously, not all technical distinctions between AI systems matter. It is unclear, for example, that administration scholars need to agonise over the distinction between recurrent neural networks \citep{sundermeyerLSTMNeuralNetworks2012} and transformers \citep{vaswaniAttentionAllYou2017}, or about whether a predictive system uses Support Vector Machines or Random Forests. Neither distinction is likely to affect governance affordances, such as how a public servant may interpret its output. Adding this technical detail may even obscure a paper's claims, or make it less clear how widely they generalise.

This paper addresses the gap these examples illustrate: the field of public administration neither has clarity on how much technical detail about AI systems is necessary, nor the tools or conventions to provide this detail. As a result, it is unclear how well past research reflects meaningful technical differences between AI systems. 

We argue that PA research would benefit greatly from a modest increase in technical precision about the studied AI systems: classifying them into a typology of just five categories based on the affordances they provide government actors. 
To that end, we make three contributions, each answering one research question:

\textbf{Which technical distinctions between AI systems matter to the study of public administration (\secref{sec:theory} -- \secref{sec:typology})?} We formulate a test: a technical distinction between two AI systems is necessary whenever their implications for public values differ, and unnecessary otherwise. Arguing that no existing AI taxonomies meet this definition, we introduce our core contribution: a technical typology of five types of AI systems, constructed such that each group has distinct public-value implications. We specify how the governance implications of each group vary across five public values.

\textbf{How well are these technical distinctions made in recent PA research (\secref{sec:analysis} -- \secref{sec:findings})?} We analyse a sample of recent highly-cited papers in public administration and digital government. We find that technical imprecision is widespread, with many papers leaving the studied system unspecified, motivating their work with systems different from the ones they study, or drawing conclusions broader than their evidence supports.

\textbf{How can researchers and policymakers ensure technical precision in the future (\secref{sec:discussion})?} We highlight common pitfalls we find in our analysis, and offer practical recommendations for future research. Specifying the system type need not be costly: we provide a handful of diagnostic questions with which studied AI systems can be placed in our typology, which can be answered using information that is usually available to researchers.
The added precision, we argue, improves both the internal validity of individual studies and the cumulative development of the field.

\section{Requirements for a Typology} \label{sec:theory}
We begin by establishing from prior literature why -- and when -- technical precision about AI systems is required, proceeding in four steps. We first formulate a test for when technical specification matters, by analysing \emph{public values}: wherever two technically different AI systems bear differently on a public value, researchers should provide enough technical precision to distinguish them. Next, we motivate the need for a novel typology by reviewing existing taxonomies of AI systems and arguing that none of them meets this requirement. Third, we describe how  \textit{affordance theory} can be used to construct a typology that does, slicing systems precisely where their governance-relevant affordances change. Finally, using the affordance lens, we specify three ways technical imprecision may weaken PA research: underspecification of studied AI systems, mischaracterisation of prior research, and overgeneralisation of conclusions.

\subsection{When Technical Precision Matters: AI and Public Values}

A large and fast-growing body of scholarship charts the impact of AI systems on government \citep{wirtzArtificialIntelligencePublic2019,valle-cruzAssessingPublicPolicycycle2020,madanAIAdoptionDiffusion2023}. Whether this work suffers from technical imprecision -- and whether such imprecision, if present, weakens its findings -- is not self-evident. To find out, we first capture systematically what the field treats as ``of interest,'' so we can then ask whether those concerns vary across technically different systems.


\emph{Public values} provide such a framework. Public values are the features of government bodies that uphold \emph{good governance} -- equity, legitimacy, and accountability among them. 
Though private institutions may exhibit some of these, public-sector organisations are distinctive in their steadfast commitment to them, and upholding them is the bedrock of public administration research. 
Given this centrality, we take all research on AI in PA to study the interaction of AI with at least one public value.

Drawing on the ``good digital governance'' framework of \citet{stalenhoefEenDialoogVoor2024}, we map key research concerns about AI in PA onto five public values (Table~\ref{tab:governance-values}). Collectively, these dimensions allow us to test the need for technical specificity: wherever two technically different systems bear on one of these values differently, enough specification to tell them apart becomes necessary. 

While these five interpretations capture a large share of the field's work, we make no claim that they are exhaustive, given the breadth of AI's potential effects on public administration. We revisit this limitation in \secref{sec:limitations}. Even so, we posit that these issues capture a substantial portion of the field's work and are therefore sufficient to demonstrate the need for greater technical precision. Adding further values or interpretations may multiply the points at which such systems diverge and may be a fruitful avenue for future research (\secref{sec:further-research}).

Below, we briefly introduce the five dimensions and their AI-relevant interpretations.

\begin{table}[ht]
\centering
\footnotesize
\renewcommand{\arraystretch}{1.1}
\begin{tabular}{@{}p{0.22\columnwidth} p{0.28\columnwidth} p{0.42\columnwidth}@{}}
\toprule
\textbf{Dimension} & \textbf{Core Value} & \textbf{AI-relevant interpretation} \\
\midrule
Democracy & Participation & \textbf{Transparency}: Decision logic must be open to public scrutiny so that citizens and representatives can inspect the basis on which authority is exercised. \\
\midrule
\multirow{2}{=}{Rule of law} & Procedural justice & \textbf{Explainability}: Consequential decisions must be reasoned and communicable to those affected, enabling meaningful contestation. \\
                              & Human rights & \textbf{Non-discrimination}: Individuals must be treated equitably regardless of protected characteristics; historical data must not encode and reproduce past inequalities. \\
\midrule
\multirow{2}{=}{Governing capability} & Quality of governance & \textbf{Implementation capacity}: The state must be able to deploy and manage AI systems in service of public purposes. \\
                                       & Responsibility & \textbf{Accountability}: Public action must be attributable to a responsible actor who can be held answerable for it. \\
\bottomrule
\end{tabular}
\caption{Good governance dimensions, core values \citep{stalenhoefEenDialoogVoor2024}, and AI-specific interpretations common in PA research.}
\label{tab:governance-values}
\end{table}

\paragraph{Participation} Public decisions should be open to scrutiny, so citizens and their representatives can inspect how authority is exercised \citep{anannySeeingKnowingLimitations2018,krollAccountableAlgorithms2017}. The AI-relevant research issue is model \textbf{transparency}. Much work evaluates whether, and how, AI systems enable or undermine public participation -- by making decision logic visible \citep{mokanderArtificialIntelligenceRationalization2024,schmitzMoralAgencyFramework2025}, opening or overwhelming new channels for public input, or embedding decisions in ways immune to public examination. 



\paragraph{Procedural Justice} Administrative law in most democratic systems mandates that consequential decisions be reasoned and communicated to those affected \citep{wachterWhyRightExplanation2017,debruijnPerilsPitfallsExplainable2022,buttaboniRegulatoryTaxonomyAI2026}. A common research theme is how the \textbf{explainability} of AI systems bears on this duty: whether a functional explanation suffices for procedural legitimacy \citep{lazarLegitimacyAuthorityDemocratic2024}, whether human-illegible decision rules leave affected citizens any meaningful way to contest a decision, and what a ``right to explanation'' can deliver when the available explanations are only post-hoc approximations.

\paragraph{Human Rights} Public bodies must treat individuals affected by their actions or decisions equitably. Scholars investigate how AI may affect such \textbf{non-discrimination}: for example, many investigate how statistical regularities in historical training data can introduce bias by encoding and reproducing past inequalities \citep{barocasBigDatasDisparate2016,corbett-daviesMeasureMismeasureFairness2023}. Another research strand is sociotechnical design: whether AI-assisted governance satisfies non-discrimination hinges substantially on how the underlying system was constructed and where bias might originate \citep{selbstFairnessAbstractionSociotechnical2019,wachterBiasPreservationMachine2021,greenFlawsPoliciesRequiring2022}.

\paragraph{Quality of Governance} We expect public organisations to be effective, efficient, and economical. Scholars frequently analyse how introducing AI requires \textbf{implementation capacity}: the organisational and technical competence to deploy and manage technology in service of public aims \citep{lawrenceBureaucraticChallengeAI2023,madanAIAdoptionDiffusion2023,neumannExploringArtificialIntelligence2024}. This includes analysis of the in-house skills to procure, integrate, maintain, and oversee AI systems, of introduced dependence on external vendors and infrastructure, and of whether deployed systems increase efficiency.

\paragraph{Responsibility} Finally, public action must be attributable to a specific public actor who can answer for it \citep{nissenbaumAccountabilityComputerizedSociety1996,bovensAnalysingAssessingAccountability2007}. An active research area is \textbf{accountability}: because AI systems distribute decision-making across complex technical architectures and lengthy supply chains, scholars ask where accountability should land -- with the official who relied on a system's output, the agency that deployed it, or the vendor that built it -- and whether existing mechanisms can still locate a responsible actor at all \citep{matthiasResponsibilityGapAscribing2004,sterzQuestEffectivenessHuman2024}.

\subsection{Why Existing AI Taxonomies are Unsuitable}

Several taxonomies of AI systems already exist, within public administration and beyond. However, we argue none of them meets the requirements of our public-value framework -- not because they are poorly constructed, but because each was built for a different purpose. 

Existing taxonomies organise AI systems using three sets of principles. First, some classify systems by their \emph{technical properties} -- such as learning paradigm, architecture, or the task completed (e.g. classification). The OECD Framework for the Classification of AI Systems, for example, maps systems along four contextual dimensions: people and organisations, technical characteristics, data and input, and task and output \citep{OECDFrameworkClassification2022}. Similarly, the INSYTE framework scores systems on eight such dimensions and renders each as a radar chart \citep{porterINSYTEClassificationFramework2025}. A second organising principle is \emph{application domain} and potentially its associated \textit{risk level}, as in the EU AI Act's tiered scheme and derived regulatory analysis \citep{lauxTrustworthyArtificialIntelligence2024,buttaboniRegulatoryTaxonomyAI2026}. Application-based classification is particularly prevalent in PA taxonomies, which catalogue systems by use case and government function \citep{berryhillHelloWorldArtificial2019,wirtzArtificialIntelligencePublic2019}. 
The third organising principle is AI systems' \emph{role in a decision}, distinguishing AI that suggests, offloads, or supersedes a human judgement \citep{roehlAutomatedAdministrativeDecisionmaking2024,konigOpportunityRenewalDisruptive2020}. 

Each of these approaches serves a distinct purpose, such as regulation \citep{buttaboniRegulatoryTaxonomyAI2026} or safety engineering \citep{porterINSYTEClassificationFramework2025}. However, it is unclear whether they result in taxonomies suitable for PA analysis. Via our public value framework, we can pose a simple criterion to evaluate this: a suitable taxonomy should group technically distinct systems that share public-value implications, and separate those whose implications differ. 

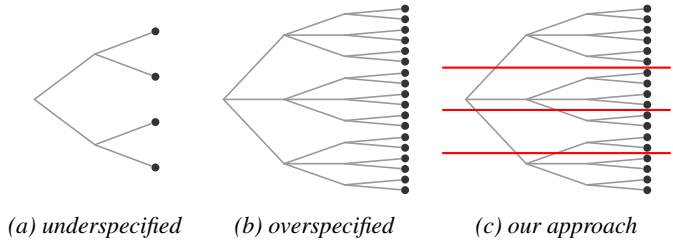
\begin{figure}[!t]
\centering

\begin{tikzpicture}[
  inode/.style = {coordinate},
  leaf/.style  = {circle, fill=black!80, minimum size=3pt, inner sep=0pt},
  ed/.style    = {draw=black!40, semithick, line cap=round},
  fault/.style = {draw=red, thick, line cap=round},
  lbl/.style   = {font=\small\itshape, text=black}
]

\node[inode] (aR) at (0,0) {};
\node[inode] (aA) at (0.8, 0.6) {};
\node[inode] (aB) at (0.8,-0.6) {};
\node[leaf] (aA1) at (1.6, 0.9) {};
\node[leaf] (aA2) at (1.6, 0.3) {};
\node[leaf] (aB1) at (1.6,-0.3) {};
\node[leaf] (aB2) at (1.6,-0.9) {};

\draw[ed] (aR)--(aA); \draw[ed] (aR)--(aB);
\draw[ed] (aA)--(aA1); \draw[ed] (aA)--(aA2);
\draw[ed] (aB)--(aB1); \draw[ed] (aB)--(aB2);

\node[lbl] at (0.8,-1.7) {(a) underspecified};

\node[inode] (bR) at (2.5,0) {};

\node[inode] (bA) at (3.3, 0.85) {};
\node[inode] (bB) at (3.3, 0.00) {};
\node[inode] (bC) at (3.3,-0.85) {};

\node[inode] (bA1) at (4.1, 1.13) {};
\node[inode] (bA2) at (4.1, 0.85) {};
\node[inode] (bA3) at (4.1, 0.57) {};
\node[inode] (bB1) at (4.1, 0.28) {};
\node[inode] (bB2) at (4.1, 0.00) {};
\node[inode] (bB3) at (4.1,-0.28) {};
\node[inode] (bC1) at (4.1,-0.57) {};
\node[inode] (bC2) at (4.1,-0.85) {};
\node[inode] (bC3) at (4.1,-1.13) {};

\node[leaf] (bA11) at (4.9, 1.20) {};
\node[leaf] (bA12) at (4.9, 1.06) {};
\node[leaf] (bA21) at (4.9, 0.92) {};
\node[leaf] (bA22) at (4.9, 0.78) {};
\node[leaf] (bA31) at (4.9, 0.64) {};
\node[leaf] (bA32) at (4.9, 0.50) {};
\node[leaf] (bB11) at (4.9, 0.35) {};
\node[leaf] (bB12) at (4.9, 0.21) {};
\node[leaf] (bB21) at (4.9, 0.07) {};
\node[leaf] (bB22) at (4.9,-0.07) {};
\node[leaf] (bB31) at (4.9,-0.21) {};
\node[leaf] (bB32) at (4.9,-0.35) {};
\node[leaf] (bC11) at (4.9,-0.50) {};
\node[leaf] (bC12) at (4.9,-0.64) {};
\node[leaf] (bC21) at (4.9,-0.78) {};
\node[leaf] (bC22) at (4.9,-0.92) {};
\node[leaf] (bC31) at (4.9,-1.06) {};
\node[leaf] (bC32) at (4.9,-1.20) {};

\draw[ed] (bR)--(bA); \draw[ed] (bR)--(bB); \draw[ed] (bR)--(bC);
\draw[ed] (bA)--(bA1); \draw[ed] (bA)--(bA2); \draw[ed] (bA)--(bA3);
\draw[ed] (bB)--(bB1); \draw[ed] (bB)--(bB2); \draw[ed] (bB)--(bB3);
\draw[ed] (bC)--(bC1); \draw[ed] (bC)--(bC2); \draw[ed] (bC)--(bC3);
\draw[ed] (bA1)--(bA11); \draw[ed] (bA1)--(bA12);
\draw[ed] (bA2)--(bA21); \draw[ed] (bA2)--(bA22);
\draw[ed] (bA3)--(bA31); \draw[ed] (bA3)--(bA32);
\draw[ed] (bB1)--(bB11); \draw[ed] (bB1)--(bB12);
\draw[ed] (bB2)--(bB21); \draw[ed] (bB2)--(bB22);
\draw[ed] (bB3)--(bB31); \draw[ed] (bB3)--(bB32);
\draw[ed] (bC1)--(bC11); \draw[ed] (bC1)--(bC12);
\draw[ed] (bC2)--(bC21); \draw[ed] (bC2)--(bC22);
\draw[ed] (bC3)--(bC31); \draw[ed] (bC3)--(bC32);

\node[lbl] at (3.7,-1.7) {(b) overspecified};

\node[inode] (cR) at (5.7,0) {};

\node[inode] (cA) at (6.5, 0.85) {};
\node[inode] (cB) at (6.5, 0.00) {};
\node[inode] (cC) at (6.5,-0.85) {};

\node[inode] (cA1) at (7.3, 1.13) {};
\node[inode] (cA2) at (7.3, 0.85) {};
\node[inode] (cA3) at (7.3, 0.57) {};
\node[inode] (cB1) at (7.3, 0.28) {};
\node[inode] (cB2) at (7.3, 0.00) {};
\node[inode] (cB3) at (7.3,-0.28) {};
\node[inode] (cC1) at (7.3,-0.57) {};
\node[inode] (cC2) at (7.3,-0.85) {};
\node[inode] (cC3) at (7.3,-1.13) {};

\node[leaf] (cA11) at (8.1, 1.20) {};
\node[leaf] (cA12) at (8.1, 1.06) {};
\node[leaf] (cA21) at (8.1, 0.92) {};
\node[leaf] (cA22) at (8.1, 0.78) {};
\node[leaf] (cA31) at (8.1, 0.64) {};
\node[leaf] (cA32) at (8.1, 0.50) {};
\node[leaf] (cB11) at (8.1, 0.35) {};
\node[leaf] (cB12) at (8.1, 0.21) {};
\node[leaf] (cB21) at (8.1, 0.07) {};
\node[leaf] (cB22) at (8.1,-0.07) {};
\node[leaf] (cB31) at (8.1,-0.21) {};
\node[leaf] (cB32) at (8.1,-0.35) {};
\node[leaf] (cC11) at (8.1,-0.50) {};
\node[leaf] (cC12) at (8.1,-0.64) {};
\node[leaf] (cC21) at (8.1,-0.78) {};
\node[leaf] (cC22) at (8.1,-0.92) {};
\node[leaf] (cC31) at (8.1,-1.06) {};
\node[leaf] (cC32) at (8.1,-1.20) {};

\draw[ed] (cR)--(cA); \draw[ed] (cR)--(cB); \draw[ed] (cR)--(cC);
\draw[ed] (cA)--(cA1); \draw[ed] (cA)--(cA2); \draw[ed] (cA)--(cA3);
\draw[ed] (cB)--(cB1); \draw[ed] (cB)--(cB2); \draw[ed] (cB)--(cB3);
\draw[ed] (cC)--(cC1); \draw[ed] (cC)--(cC2); \draw[ed] (cC)--(cC3);
\draw[ed] (cA1)--(cA11); \draw[ed] (cA1)--(cA12);
\draw[ed] (cA2)--(cA21); \draw[ed] (cA2)--(cA22);
\draw[ed] (cA3)--(cA31); \draw[ed] (cA3)--(cA32);
\draw[ed] (cB1)--(cB11); \draw[ed] (cB1)--(cB12);
\draw[ed] (cB2)--(cB21); \draw[ed] (cB2)--(cB22);
\draw[ed] (cB3)--(cB31); \draw[ed] (cB3)--(cB32);
\draw[ed] (cC1)--(cC11); \draw[ed] (cC1)--(cC12);
\draw[ed] (cC2)--(cC21); \draw[ed] (cC2)--(cC22);
\draw[ed] (cC3)--(cC31); \draw[ed] (cC3)--(cC32);

\draw[fault] (5.4, 0.42) -- (8.4, 0.42);
\draw[fault] (5.4, -0.14) -- (8.4, -0.14);
\draw[fault] (5.4, -0.71) -- (8.4, -0.71);

\node[lbl] at (6.9,-1.7) {(c) our approach};

\end{tikzpicture}
\caption{Conceptual diagram of three approaches to classifying AI systems. (a)~An \emph{underspecified} taxonomy fails to distinguish between systems with different governance implications. (b)~An \emph{overspecified} taxonomy draws too many distinctions between systems, obscuring differences in their affordances. (c)~Rather than generating a novel taxonomy from scratch, \emph{our approach} identifies PA-relevant affordance thresholds (solid red lines) within existing taxonomies.}
\label{fig:taxonomy-comparison}

\end{figure}

Existing taxonomies fail this test in three ways, two of which are sketched in Figure~\ref{fig:taxonomy-comparison}. First, many \emph{underspecify}: they group systems whose governance-relevant properties differ. A shared risk tier or use-case label can place a predictive policing model beside a hospital triage assistant, though the two diverge in explainability, bias mechanisms, and accountability. Second, especially technical taxonomies \emph{overspecify}: they split systems too granularly, even where their public-value implications match. Crucially, having too many categories makes it unclear where meaningful distinctions lie. Specifying that a studied model is a random forest, for example, does not clarify how takeaways may transfer to neural networks. 
Third, many \emph{conflate functional and technical categories}. A ``suggesting'' system, for example, could produce a single risk score or paragraphs of text, which evidently vary in governance-relevant dimensions. 

Given that none of the mentioned taxonomies were designed for PA analysis, it is understandable that none pass these tests. However, there is therefore a clear need for a typology of AI systems targeted at clarifying their public-value implications. Currently, researchers risk either underspecifying their scope by using no taxonomy at all -- invoking ``AI'', ``algorithms'', or ``automated decision-making'' generically -- or relying on an unsuitable existing taxonomy, which does not clearly track such properties.

\subsection{Affordance Theory as the Organising Principle}

Affordance theory provides a suitable organising principle for a better-calibrated taxonomy. Developed by Gibson \citeyearpar{gibsonEcologicalApproachVisual1979} and elaborated in organisational and information-systems research \citep{zammutoInformationTechnologyChanging2007, majchrzakTechnologyAffordancesConstraints2013, leonardiWhenFlexibleRoutines2011}, affordances describe what a technology makes possible or forecloses for a particular actor in a particular setting.  For example, \citet{bovensStreetLevelSystemLevelBureaucracies2002} famously motivate the concept of screen-level bureaucracy with the affordances of a digital form over a paper one.

Affordance theory reflects that a technology's public-value impacts depend on its architecture -- but are not solely defined by it. They are also shaped by the interaction between the technology and the goals, capacities, and context of those who use it. Identical algorithms can produce different organisational outcomes depending on the setting \citep{meijerAlgorithmizationBureaucraticOrganizations2021}, and how AI systems affect public values is a question of socio-technical design \citep{schmitzMoralAgencyFramework2025}.

Indeed, such organisational and contextual analysis remains vital. Throughout this work, we do not suggest that technical detail should \textit{replace} such analysis, but that it must \textit{complement} it: the technical architecture defines which affordances exist to begin with. For example, open-weights large language models (LLMs) can afford public organisations the processing of sensitive data where proprietary models do not, but this need not imply that they are used for that purpose \citep{robinsonOpenOpensourceAI2026}.

An affordance-based typology is therefore suitable to bridge organisational and technical analysis. In \secref{sec:typology}, we construct this typology by drawing boundaries we term \emph{affordance thresholds}: distinctions \textit{only} between sets of AI systems with different public value-relevant affordances, as visualised in Fig. \ref{fig:taxonomy-comparison} (c). A technical difference between two systems that alters whether a citizen can contest a decision, or whether an auditor can inspect its logic, produces such a threshold: the typology should distinguish between them. In contrast, a shift that does not affect what actors can do -- for example, one that only improves predictive accuracy -- does not cross an affordance threshold, and the systems should remain in the same category.


\begin{figure*}[t]
    \centering

\colorlet{subsetc}{blue!55!black}
\colorlet{thrc}{orange!85!black}

\begin{tikzpicture}[
    font=\footnotesize,
    leafbox/.style = {draw=black!55, rounded corners=2pt, fill=black!4,
                      minimum height=7mm, inner xsep=6pt, anchor=west},
    thr/.style     = {black, line width=0.7pt, line cap=round, line join=round},
    vthr/.style    = {thrc, line width=1.0pt, line cap=round, line join=round},
    vsub/.style    = {subsetc, line width=1.0pt, line cap=round, line join=round},
    sglyph/.style  = {subsetc, fill=white, inner sep=1pt},
    hd/.style      = {anchor=west, font=\footnotesize\bfseries},
    cts/.style     = {anchor=west, align=left, text width=3.4cm},
    dr/.style      = {anchor=west, align=left, text width=3.1cm, font=\footnotesize\itshape},
    ex/.style      = {anchor=west, align=left, text width=3.6cm},
    rule/.style    = {black, line width=0.6pt},
    srule/.style   = {black!18, line width=0.4pt},
  ]

  \def\xleaf{3.2}  \def\xcts{6.0}  \def\xdr{9.7}  \def\xex{13.1}
  \def\xL{0.15}    \def\xR{16.8}

  \draw[rule] (\xL,6.95) -- (\xR,6.95);                       
  \node[hd] at (\xleaf,6.55) {System Type};
  \node[hd] at (\xcts,6.55)  {Technical Description};
  \node[hd] at (\xdr,6.55)   {Diagnostic Question};
  \node[hd] at (\xex,6.55)   {Illustrative example};
  \draw[rule] (\xL,6.15) -- (\xR,6.15);                       
  \draw[srule] (2.95,4.9) -- (\xR,4.9);                       
  \draw[srule] (2.95,3.5) -- (\xR,3.5);
  \draw[srule] (2.95,2.1) -- (\xR,2.1);
  \draw[srule] (2.95,0.7) -- (\xR,0.7);

  \draw[thr] (0.15,4.55) -- (0.7,4.55);

  \draw[thr]  (0.7,5.6)  -- (\xleaf,5.6);                     
  \draw[thr]  (0.7,3.5)  -- (1.4,3.5);                        
  \draw[thr]  (1.4,4.2)  -- (\xleaf,4.2);                     
  \draw[thr]  (1.4,2.8)  -- (\xleaf,2.8);                     
  \draw[vthr] (0.7,3.5)  -- (0.7,5.6);                        
  \draw[vthr] (1.4,2.8)  -- (1.4,4.2);                        
  \fill[thrc] (0.7,4.66) -- (0.81,4.55) -- (0.7,4.44) -- (0.59,4.55) -- cycle;
  \fill[thrc] (1.4,3.61) -- (1.51,3.50) -- (1.4,3.39) -- (1.29,3.50) -- cycle;
  \node[inner sep=0pt] (C1) at (0.7,4.55) {};
  \node[inner sep=0pt] (C2) at (1.4,3.50) {};

  \draw[vsub] (2.0,2.8) -- (2.0,1.4);                         
  \draw[thr]  (2.0,1.4) -- (\xleaf,1.4);                      
  \draw[vsub] (2.6,1.4) -- (2.6,0.0);                         
  \draw[thr]  (2.6,0.0) -- (\xleaf,0.0);                      
  \node[sglyph] (SB) at (2.0,2.10) {$\subset$};
  \node[sglyph] (SG) at (2.6,0.70) {$\subset$};

  \node[leafbox] at (\xleaf,5.6) {\textbf{Hand-coded}};
  \node[leafbox] at (\xleaf,4.2) {\textbf{Glass-box}};
  \node[leafbox] at (\xleaf,2.8) {\textbf{Black-box}};
  \node[leafbox] at (\xleaf,1.4) {\textbf{General-purpose}};
  \node[leafbox] at (\xleaf,0.0) {\textbf{Agentic}};

  \node[cts] at (\xcts,5.6) {Discretion or policy encoded as explicit rules \citep{enqvistRulebasedAIdrivenBenefits2024}.};
  \node[cts] at (\xcts,4.2) {Rules learned from data, but legible to experts \citep{rudinStopExplainingBlack2019}.};
  \node[cts] at (\xcts,2.8) {Learned logic defies expert inspection \citep{burrellHowMachineThinks2016}.};
  \node[cts] at (\xcts,1.4) {General, externally pre-trained models \citep{bommasaniOpportunitiesRisksFoundation2021}.};
  \node[cts] at (\xcts,0.0) {General-purpose model scaffolded to act over time, e.g. via tool-use \citep{yaoReActSynergizingReasoning2023}.};

  \node[dr] at (\xdr,4.2) {Are the rules the system follows learned from data, rather than authored in code?};
  \node[dr] at (\xdr,2.8) {Is it infeasible for any human expert to understand the system's learned logic and internal functioning?};
  \node[dr] at (\xdr,1.4) {Does the system adapt a generally-trained model to a specific task?};
  \node[dr] at (\xdr,0.0) {Is the general-purpose model scaffolded to act autonomously over time, e.g. via tools?};

  \node[ex] at (\xex,5.6) {Florida ACCESS benefits platform \citep{citronTechnologicalDueProcess2008}.};
  \node[ex] at (\xex,4.2) {Allegheny Family Screening Tool \citep{vaithianathanDevelopingPredictiveModels2017}.};
  \node[ex] at (\xex,2.8) {COMPAS recidivism scoring \citep{dresselAccuracyFairnessLimits2018}.};
  \node[ex] at (\xex,1.4) {GPT-4 via Azure for citizen queries \citep{brightGenerativeAIAlready2025}.};
  \node[ex] at (\xex,0.0) {Benefits agent querying registries autonomously \citep{ilvesAgenticStateRethinking2025}.};

  \draw[rule] (\xL,-0.7) -- (\xR,-0.7);
  \fill[thrc] (0.40,-0.94) -- (0.51,-1.05) -- (0.40,-1.16) -- (0.29,-1.05) -- cycle;
  \node[anchor=west, text=black!75] at (0.7,-1.05)
        {\textit{Threshold}: qualitative distinction};
  \node[sglyph] at (0.40,-1.55) {$\subset$};
  \node[anchor=west, text=black!75] at (0.7,-1.55)
        {\textit{Subset}: special case of the type above};

\end{tikzpicture}
    \caption{\textbf{Overview of Technical Typology of AI Systems.} The five system classes (left) are separated by two kinds of relationships. Orange diamonds mark \textit{thresholds} -- qualitative distinctions; $\subset$ markers mark \textit{subsets}, where a system is a special case of the one above. Diagnostic questions distinguish each class from the class above, and can therefore be used as a specification tool, as described in \secref{sec:recommendations}.}
    \label{fig:typology}
\end{figure*}

\subsection{Three Forms of Technical Imprecision} \label{sec:failures}

Using the lens of affordance thresholds, we can phrase more precisely how technical imprecision on AI could weaken PA research. We theorise three potential forms of imprecision here;  in our review of the field (\secref{sec:analysis}), we operationalise these definitions and measure how often each occurs.

\begin{itemize}
    \item \textit{Underspecification} could arise when too little technical detail is provided, such that it is not clear what the affordance profile of a studied AI system is. For example, describing a system only as ``AI'' or an ``algorithm'' may not allow its specific affordances to be recovered, such that the generalisability of any claims made cannot be verified.
    \item \textit{Mischaracterisation} may arise when a paper motivates its approach with one type of system, then studies another. For example, a paper that opens on the dangers of opaque, black-box risk scoring and then examines a rule-based eligibility calculator ports over inaccurate assumptions about the system's afforded transparency.
    \item \textit{Overgeneralisation} could occur if a paper presents its findings as more general than the evidence supports. For example, a conclusion about ``AI in government'' drawn from the study of a black-box model may not apply to general-purpose models.
\end{itemize}

Evidently, claims can cross affordance thresholds without being imprecise in one of these ways: a well-scoped insight from a case study of one system could readily generalise to many other types, for example. Technical precision allows us to distinguish which claims do so validly.

\section{Typology of AI Systems in Public Administration} \label{sec:typology}



This section introduces our core contribution: a technical typology of five classes of AI systems, presented in Figure \ref{fig:typology}.
We define and describe each class of system, specify when a system crosses the ``affordance threshold'' between classes, and describe the distinct public-value implications of each class \citep{stalenhoefEenDialoogVoor2024}. These are summarized in Table \ref{tab:ai_layers_governance}.

We deliberately do not attempt to define ``AI'' ourselves. All of the levels in our typology have been termed ``AI'' by some widely-cited papers, as we analyse in \secref{sec:pitfalls}. Rather, we provide a simple but nuanced vocabulary for talking about these types of systems.




\subsection{Hand-coded systems}
The first layer is \emph{hand-coded systems}: systems whose decision rules are authored in code rather than learned from data. The canonical example is rule-based public benefit administration \citep{enqvistRulebasedAIdrivenBenefits2024}, and the category has been extensively studied in the e-governance and digital government literatures \citep{dunleavyDigitalEraGovernance2006,zouridisAutomatedDiscretion2020}.

Hand-coded systems drive the shift from street-level to system-level bureaucracy \citep{bovensStreetLevelSystemLevelBureaucracies2002}, in which discretion is encoded in software rather than exercised case-by-case by individual officials.

\paragraph{Affordance: efficiency and traceability}
While hand-coded systems can improve efficiency, traceability, and standardise procedures, they simultaneously diffuse accountability \citep{citronTechnologicalDueProcess2008} and flatten local complexities, exacerbating existing legibility dynamics \citep{scottSeeingStateHow1998}. Furthermore, the shift to digital systems enables large-scale data collection and analysis, with significant implications for privacy \citep{zuboffBigOtherSurveillance2015}.

\paragraph{Challenge: complexity and bias from data-rule interaction}
It is important to note that rule complexity can be immense even within hand-coded systems. While the logic is authored rather than learned, thousands of intersecting rules can still exceed human cognitive limits \citep{simonAdministrativeBehavior1947}. Moreover, understanding a rule does not equate to understanding its interaction with real-world data. Even fully legible, hand-coded rules can introduce bias because discrimination is a function of how the system affects outcomes in practice \citep{wachterWhyFairnessCannot2021}. For instance, a seemingly neutral hand-coded rule that declines benefits if an applicant has a continuous \texttt{unemployed\_duration > 6 months} may inadvertently discriminate against women taking maternity leave. Thus, bias and discrimination can manifest through the interaction between fixed rules and contextual realities, entirely independent of statistical learning.

\paragraph{Boundary}
Whether hand-coded systems truly fall under ``AI'' is contested. Some scholars explicitly include rule-based systems \citep{seltenJustThoughtStreetlevel2023}, and scholarship on Robotic Process Automation (RPA) often uses the language of AI. Our aim is not to settle that question, but to sharpen the vocabulary used to distinguish between different kinds of systems. Hand-coded systems remain in this layer so long as their rules are authored in code; once rules begin to be derived from data, the system crosses into the glass-box layer.

\exampleHardCoded

\begin{table*}[!htbp]
\centering
\footnotesize
\renewcommand{\arraystretch}{1.1}
\begin{tabular}{@{}p{2.4cm} p{4.4cm} p{4.4cm} p{4.4cm}@{}}
\toprule
\textbf{Layer} & \textbf{Democracy} \newline \textit{(Participation)} & \textbf{Rule of Law} \newline \textit{(Justice \& Rights)} & \textbf{Governing Capability} \newline \textit{(Quality \& Responsibility)} \\
\midrule

\textbf{Hand-coded}
& Shifts to system-level administration, flattening local complexities \citep{bovensStreetLevelSystemLevelBureaucracies2002, scottSeeingStateHow1998}.
& Standardises procedures but expands data matching, risking privacy \citep{citronTechnologicalDueProcess2008}. 
& Enhances efficiency and traceability, yet diffuses accountability \citep{bovensStreetLevelSystemLevelBureaucracies2002, citronTechnologicalDueProcess2008}. \\
\midrule

\textbf{Glass-box}
& Adapts to case variation but risks reproducing inequalities and displacing public values \citep{dignazioDataFeminism2023, wachterBiasPreservationMachine2021, greenAlgorithmicRiskAssessments2021}.
& Legibility permits auditing of learned features, though accountability blurs \citep{sandvigAuditingAlgorithmsResearch2014}. 
& Shared decision-making among model, data, and user creates ``moral crumple zones'' \citep{elishMoralCrumpleZones2019}. \\
\midrule

\textbf{Black-box}
& Boosts performance in unstructured domains but restricts citizen capacity to contest decisions \citep{wangWhatTypeAlgorithm2023, valle-cruzExploringNegativeImpacts2024}.
& Threatens procedural justice via justification deficits, impeding legal verification \citep{grimmelikhuijsenLegitimacyAlgorithmicDecisionmaking2022, rudinStopExplainingBlack2019}. 
& Inscrutiable internal logic obscures decision pathways and contests responsibility \citep{cobbeUnderstandingAccountabilityAlgorithmic2023, janssenDataGovernanceOrganizing2020}. \\
\midrule

\textbf{General-purpose}
& Lowers adoption barriers via natural language, but democratisation of control remains partial \citep{brightGenerativeAIAlready2025, hashemMappingPotentialGenerative2025}.
& Persuasiveness and unfaithful explanations risk automation bias; inaccessible training data heightens privacy risks \citep{maynePositiveCaseFaithfulness2026, benderDangersStochasticParrots2021}. 
& Distributed responsibility and hardware dependence create vendor lock-in and complicate auditing \citep{brownAllocatingAccountabilityAI2023, mokanderAuditingLargeLanguage2024} \\
\midrule

\textbf{Agentic}
& Reduces citizen friction but scales demand, risking administrative overload \citep{ilvesAgenticStateRethinking2025, yunImprovingCitizengovernmentInteractions2024, marquesLeveragingLLMsStreamline2025}. 
& Diffuses discretionary power by acting proactively, challenging traditional procedural safeguards and accountability \citep{chanHarmsIncreasinglyAgentic2023}. 
& Demands runtime oversight rather than ex post evaluation; capabilities present a ``jagged frontier'' \citep{schmitzOversightStructuresAgentic2025, chanVisibilityAIAgents2024, dellacquaNavigatingJaggedTechnological2026}. \\
\bottomrule
\end{tabular}
\caption{Summary of public-value implications of each layer of the taxonomy.}
\label{tab:ai_layers_governance}
\vspace{-4mm}
\end{table*}

\subsection{Glass-box systems}

The second layer is \emph{glass-box systems}: systems whose decisions rules are learned from data, but whose learned logic remains inspectable -- at least by experts \citep{burrellHowMachineThinks2016,rudinStopExplainingBlack2019}.  Canonical methods include linear and logistic regression \citep{foxAppliedRegressionAnalysis2015}, decision trees \citep{devilleDecisionTrees2013}, Principal Component Analysis \citep{abdiPrincipalComponentAnalysis2010}, and TF-IDF \citep{bafnaDocumentClusteringTFIDF2016}  --  methods used for prediction, classification, and representation across applications such as child-welfare risk modelling \citep{hallSystematicReviewSophisticated2024}, welfare-recipient predictions \citep{sansoneUsingMachineLearning2023}, and policy-document text analysis \citep{altaweelDocumentsDataContent2019}.

This is a binary shift into machine learning; rules are no longer specified exhaustively but derived from data. The shift opens up case-level adaptation and leveraging of historical data patterns, while introducing new governance challenges around bias and accountability.

\paragraph{Affordance: case-level adaptation}
While glass-box systems provide novel affordances  --  glass-box systems can adapt to historical variations and improve administrative fit where fixed rules are too coarse  --  learned rules are deeply entangled with their training data. They reproduce historical inequalities and encode patterns that do not reflect current public values or legal commitments \citep{dignazioDataFeminism2023,wachterBiasPreservationMachine2021}. This is especially problematic in public administration, since predictive accuracy is not itself the goal; decisions must also reflect normative commitments \citep{grimmelikhuijsenLegitimacyAlgorithmicDecisionmaking2022}. Decision-makers therefore risk placing undue weight on statistical outputs, even when they displace other relevant public values \citep{greenAlgorithmicRiskAssessments2021}.

\paragraph{Challenges: transparency-fairness tradeoff and responsibility}
Importantly, transparency and rule complexity exist on a spectrum. While the logic of a glass-box system is theoretically inspectable, a linear regression with thousands of variables or a highly branched decision tree can easily exceed human cognitive limits \citep{liptonMythosModelInterpretability2018}. Furthermore, transparency is often antagonistic with respect to fairness. Because glass-box systems are constrained in their complexity, the easiest way to achieve acceptable baseline performance is often to optimise for the majority while ignoring minority groups or complex edge cases \citep{ferryTamingTriangleInterplays2025}. This flexibility-interpretability tradeoff is a technical inevitability -- and a key factor in explaining why public administrators might choose to implement more complex, black-box systems.

At the same time, responsibility becomes more diffuse: when a rule is generated from data, accountability can blur between model, data, designer, and user, creating the conditions for ``moral crumple zones'' \citep{elishMoralCrumpleZones2019}. Still, compared to later black-box systems, glass-box systems remain relatively auditable because scrutiny can occur at the level of learned rules and features rather than relying primarily on indirect experiments \citep{sandvigAuditingAlgorithmsResearch2014}.

\paragraph{Boundary}
Glass-box systems remain in this layer so long as their learned logic is meaningfully inspectable; once that ceases to be the case, they move into the black-box layer.

\exampleTransparent

\subsection{Black-box systems}
The third layer is \emph{black-box systems}: systems whose learned logic resists meaningful inspection, even by experts \citep[systems with ``algorithmic opacity'' following][]{burrellHowMachineThinks2016}. Canonical examples include deep neural networks \citep{goodfellowDeepLearning2016}, random forests \citep{breimanRandomForests2001}, and UMAP \citep{mcinnesUMAPUniformManifold2020} deployed across unstructured domains \citep[e.g., face recognition;][]{rezendeFacialRecognitionPolice2020} and accuracy critical applications \citep[e.g., medical imaging;][]{rystromOxEnsembleFairEnsembles2026}.

The exact threshold between glass-box and black-box systems lies on a spectrum of algorithmic complexity: as models grow more complex, expert inspection becomes algorithmically infeasible. Consequently, approximate post-hoc explanation becomes a technical necessity \citep{mittelstadtExplainingExplanationsAI2019}, and audits must become indirect through behavioural testing \citep{sandvigAuditingAlgorithmsResearch2014}  --  eroding the ability to justify, contest, or mechanistically explain individual decisions.

\paragraph{Affordance: complex performance}
What black-box systems lack in transparency, they compensate for through improved performance in high-dimensional or unstructured domains, where simpler glass-box models struggle \citep{krizhevskyImageNetClassificationDeep2012}. As a result, black-box systems are widely deployed in government despite their governance challenges \citep{valle-cruzExploringNegativeImpacts2024}.

\paragraph{Challenges: procedural justice and responsibility}
Opacity specifically threatens procedural justice. First, it creates a justification problem: it becomes difficult to ensure legal requirements are consistently met when the decision-making logic cannot be inspected \citep{grimmelikhuijsenLegitimacyAlgorithmicDecisionmaking2022}. Second, it creates a contestation problem: citizens lack the knowledge or mechanisms to challenge decisions effectively \citep{wangWhatTypeAlgorithm2023}. Third, it creates an explanation problem: even where a ``right to explanation'' is invoked, explanations are limited to local approximations rather than exact causal pathways \citep{wachterCounterfactualExplanationsOpening2017}.

Opacity also makes responsibility more contested. Complex supply chains of models and data make decision pathways less reconstructible \citep{cobbeUnderstandingAccountabilityAlgorithmic2023}. Still, compared to later systems, black-box models are typically developed and trained within organisational boundaries, meaning that data collection and model development remain under institutional control.

\paragraph{Boundary}
Black-box systems remain in this layer so long as the underlying model is trained directly for the task at hand; once the model is instead pre-trained for a general objective and adapted to downstream tasks, the system crosses into the general-purpose layer.

\exampleOpaque

\subsection{General-purpose systems} \label{sec:typology:general}
The fourth layer is \emph{general-purpose systems}: systems pre-trained on general tasks -- such as next-token prediction -- that can be adapted to diverse downstream applications through mechanisms like transfer learning or natural language instructions \citep{brownLanguageModelsAre2020}. 
Canonical examples include large language models (LLMs) like ChatGPT \citep{openaiChatGPTOptimizingLanguage2022}, various embedding models \citep{devlinBERTPretrainingDeep2019}, and vision models  --  all of which are being increasingly deployed by public sector organisations \citep{brightGenerativeAIAlready2025}. 

General-purpose systems are a subset of black-box systems, distinguished from other black-box systems by broad pre-training for general capabilities \citep{bommasaniOpportunitiesRisksFoundation2021}, rather than narrow, task-specific utilisation of experience \citep{mitchellMachineLearning2013}.  
Consequently, the primary shift is one of \emph{externalisation}. Because of the complexity of general-purpose systems, the training process, data provenance, and model design is often done by external model providers and thus no longer fully inspectable or controllable by the deploying organisation \citep{mokanderAuditingLargeLanguage2024}. 

\paragraph{Affordance: lower adoption barriers}
A key affordance of externalisation is that it lowers barriers to adoption. Because much of the technical complexity sits outside the organisation, general-purpose systems can be integrated into administrative processes without specialised AI engineering expertise \citep{brightGenerativeAIAlready2025}. In addition, natural-language interfaces can make existing bureaucratic systems more accessible by mediating interactions between citizens and administrative procedures \citep{hashemMappingPotentialGenerative2025}. However, this ``democratisation'' is partial: while it becomes easier to use such systems, control over their behaviour remains limited. This is reinforced by their computational requirements. Many general-purpose systems require specialised hardware and are therefore deployed via cloud infrastructure, creating further dependence on external providers and reinforcing the externalisation of control \citep{qiuLockinHypothesisStagnation2025}. 

\paragraph{Challenges: distributed responsibility and procedural justice}
This transition fundamentally alters how the system utilises experience to improve performance. While traditional machine learning uses experience to improve at a specific, narrow task, general-purpose systems leverage broad pre-training to develop versatile capabilities that are then applied to specific downstream contexts \citep{brownLanguageModelsAre2020}. In technical research, these are often referred to as `foundation models' \citep{bommasaniOpportunitiesRisksFoundation2021}, though the term general-purpose highlights their functional role in the public sector as adaptable building blocks. By shifting the technical burden of training to external providers, these systems introduce a critical governance vulnerability: the externalisation of evaluation. Organisations may be tempted to rely on a provider’s general benchmarks rather than rigorously testing the system for the specific, local context of an administrative task \citep{rystromAgentBenchmarksFail2026}.

Externalisation primarily threatens responsibility and accountability. Responsibility is not only diffused but distributed across a complex supply chain involving model developers, platform providers, and deploying organisations, each with only partial control over outcomes \citep{brownAllocatingAccountabilityAI2023}. This complicates auditing; understanding a given application requires evaluating not just the downstream implementation, but also the underlying model and the opaque governance practices of its provider \citep{mokanderAuditingLargeLanguage2024}. In practice, this makes it difficult to determine whether problematic outcomes, such as discriminatory bias, stem from the training data, the model design, or the specific prompting patterns. Furthermore, the scale and opacity of training data intensify human rights concerns regarding privacy and the automated reproduction of harmful social patterns \citep{benderDangersStochasticParrots2021}.

Finally, general-purpose systems pose significant challenges for procedural justice. Because many applications rely on natural-language interfaces, there is a temptation to treat the system itself as a source of explanation \citep{zhuExplanationEraLarge2024}. Yet, such linguistic outputs are not guaranteed to reflect the underlying basis of a decision \citep{maynePositiveCaseFaithfulness2026}. The persuasive and anthropomorphic nature of these outputs can make explanations appear authoritative even when they are unfaithful \citep{salviConversationalPersuasivenessGPT42025}, potentially increasing automation bias and reinforcing the ``moral crumple zones'' that obscure human responsibility \citep{elishMoralCrumpleZones2019}.

\paragraph{Boundary}
While general-purpose systems are versatile, they remain primarily reactive, mapping inputs to outputs. Once systems move beyond this paradigm to act proactively and interact with their environment over time, they enter the final layer of the typology.

\exampleFoundation

\subsection{Agentic systems} \label{sec:typology:agentic}
The final layer of the typology is \emph{agentic systems}: AI systems that can pursue complex and general goals, act with autonomy, and affect their environment \citep{kasirzadehCharacterizingAIAgents2025}. Examples include large language models with access to external `tools' and APIs \citep{yaoReActSynergizingReasoning2023} as well as autonomous vehicles \citep[e.g.,][]{raoDeepLearningSelfdriving2018}.

Agentic systems are a subset of general-purpose systems, as they are usually created by ``scaffolding'' or augmenting general-purpose systems. The conceptual shift is therefore from strictly reactive input-output mapping to proactive, iterative loops of reasoning and action within open-ended environments \citep{yaoReActSynergizingReasoning2023}. As a result, the primary governance shift is from instance-level decision-making (e.g., the correctness of a classification) to \emph{process-level action} (e.g., the suitability of guardrails for an agent), requiring a move from ex-post evaluation towards runtime monitoring and intervention in ongoing system behaviour \citep{schmitzOversightStructuresAgentic2025}. 

\paragraph{Affordance: service delivery and citizen interface}
This transition represents a profound shift in the evolving definition of the machine learning ``task.'' Rather than producing discrete, static outputs, agentic systems use experience to navigate sequences of steps over time. In the public sector, this promises to transform service delivery by bridging fragmented infrastructures; as argued by \citet{ilvesAgenticStateRethinking2025}, agentic systems can act across heterogeneous data sources and administrative systems to streamline bureaucratic procedures. From the citizen's perspective, agentic systems also reshape the interface with government by acting as proactive intermediaries. By helping citizens navigate complex eligibility requirements and administrative hurdles, these agents can significantly reduce the effort required to access public services \citep{yunImprovingCitizengovernmentInteractions2024,joAITrustReshaping2025}.

\paragraph{Challenges: jagged reliability, surging demand, and runtime oversight}
The reliability of these systems in public-sector environments remains a significant concern. There are currently no evaluations that accurately capture their capacity for administrative tasks \citep{rystromAgentBenchmarksFail2026}, a problem compounded by the ``jagged frontier'' of agentic capabilities, which makes it difficult to predict which tasks they will perform reliably and where they will fail \citep{dellacquaNavigatingJaggedTechnological2026}.

Furthermore, the reduction in interaction costs may substantially increase the total demand for public services, as agents can interface with government systems at scale on behalf of individuals \citep{marquesLeveragingLLMsStreamline2025}. Without appropriate institutional countermeasures, this surge in automated requests may challenge the fundamental processing capacity and responsiveness of administrative systems.

Finally, agentic systems introduce substantial risks for responsibility and procedural justice \citep{chanHarmsIncreasinglyAgentic2023}. Because these systems act autonomously across time, responsibility often shifts away from discrete moments of decision toward ongoing processes, making it harder to attribute specific outcomes to individual actors \citep{schmitzMoralAgencyFramework2025}. Governance therefore requires new capacities for runtime oversight -- the ability to monitor, constrain, and intervene in live system behaviour as it unfolds \citep{chanVisibilityAIAgents2024}. Without such mechanisms, the use of agentic systems risks diffusing discretionary power away from human decision-makers and undermining established structures of accountability. Ultimately, agentic systems mark a definitive shift from systems that produce outputs to systems that act, introducing a distinct set of governance challenges that cannot be addressed through existing static approaches alone \citep{chanHarmsIncreasinglyAgentic2023}.

\exampleAgentic

\section{Analysing Technical Imprecision in Public Administration Research on AI} \label{sec:analysis}

We next validate our typology by analysing whether it would improve the precision of existing research. 
To do so, we code impactful public administration and digital government papers on AI from the last seven years, evaluating whether technical specification using the typology would mitigate their imprecision (\secref{sec:failures}).
This section details our methodology. We present the findings in \secref{sec:findings} and discuss their significance in \secref{sec:discussion}.


\subsection{Data}

To identify impactful papers studying AI in government and public administration, we conduct a systematic literature search using OpenAlex \citep{priemOpenAlexFullyopenIndex2022}. We focus on leading journals in public administration and digital governance, selected based on their relevance and citation impact within the field \citep{thielResearchMethodsPublic2021,heeksAnalyzingEgovernmentResearch2007}. The full venue list and keyword query are reported in \ref{app:data}; 
all data, code, and materials are available online.\footnote{\url{https://anonymous.4open.science/r/AITypology4PA-4FC6/}}

We include papers published between 2019 and 2025. This period ensures coverage of all categories in our typology, with a slight underrepresentation of agentic systems, which only began emerging in 2023 \citep{yaoReActSynergizingReasoning2023}. From this pool, we select the most-cited papers per year separately for public administration and digital government venues. We use citations as a proxy for impact -- a common but contested heuristic \citep{flyvbjergClassicsExistMegaproject2018} -- and sample per year to mitigate temporal bias in citation accumulation \citep{bornmannWhatCitationCounts2008}.

We determine how many papers to include per venue type and year through a pre-specified stability analysis of the aggregate estimates produced by our sampling rule. We evaluate whether the three main outcomes remain stable as additional papers are added within each stream-year cell. This follows the logic of stability-based sample adequacy, in which estimates are considered sufficient once they remain within a prespecified tolerance corridor as samples are added \citep{schonbrodtWhatSampleSize2013}, while also drawing on work on incremental-sampling thresholds \citep{guestSimpleMethodAssess2020}. We iteratively expand the corpus, recompute the aggregate outcomes, and estimate uncertainty via bootstrap resampling \citep{davisonBootstrapMethodsTheir1997}. Full methodological details are provided in \ref{app:saturation}. We find that $K=8$ papers per venue type (digital government and public administration) is sufficient to stabilise the aggregate outcomes under our sampling rule, yielding an initial set of 109 papers.

We then manually screen all papers and exclude those that use AI purely as a methodological tool \citep[e.g., using AI to predict corruption;][]{limaPredictingExplainingCorruption2020} or that do not treat AI as an empirical or theoretical subject. The final corpus consists of 91 papers. The full screening flow, including counts at each stage, is reported in \ref{app:data} (Figure~\ref{fig:prisma_flow}).

\subsection{Coding}
Once we have selected high-impact papers, we code how each paper specifies, motivates, and generalises about AI systems. We conduct structured qualitative coding, following established procedures for systematic content analysis \citep{saldanaCodingManualQualitative2025,krippendorffContentAnalysisIntroduction2019}. 

\paragraph{Coding Scheme} The primary unit of coding is not the entire paper, but a \emph{strand}: a concise summary of a key claim in the paper. An example strand is ``public organisations are not held to a lower responsibility standard for algorithmic versus human discrimination''. This meso-level analysis \citep{milesQualitativeDataAnalysis2014} has two advantages: it captures papers with more nuance, and allows systematic comparison of claims within individual papers.

For each strand, we code three pieces of metadata. First, we code it as either empirical, motivation, or conclusion. \textit{Empirical} strands identify which AI systems or class of systems the paper studies; \textit{motivation} strands summarise how the paper positions itself in existing AI literature; and \textit{conclusion} strands summarise the core claims the paper makes. Second, coders classify the AI system the strand addresses using our typology (\secref{sec:typology}). Where the paper provides insufficient technical detail to determine the system type, the strand is coded as \emph{underspecified}. Third, coders assign one or more public-value dimensions from the governance framework \citep{stalenhoefEenDialoogVoor2024}. 

Conceptual papers and literature reviews are coded using the same procedure. Where a paper studies no concrete system, coders treat the paper's main motivating examples or conceptualisation of `AI' as the empirical strands. If this conceptualisation provides enough information to identify a system type, it is classified as such; if it deliberately identifies a broad category within an affordance threshold, it is coded as \emph{justifiably generic}; otherwise, it is coded as \emph{underspecified}. 

For analysis, the strands are aggregated on the paper level as described in \secref{sec:operationalisation}. This structure enables systematic comparison between the systems used to motivate a paper, the systems actually studied, and the systems to which conclusions are applied.

\subsection{Operationalising Imprecision}
\label{sec:operationalisation}

Using the \textit{strand} construct, we formalise the three failures introduced in \secref{sec:failures}. We treat each as paper-level outcomes, measured by aggregating strand-level codes. Where relevant, we apply an \emph{any-mismatch} rule: a paper is flagged if at least one strand exhibits the imprecision. This reflects our interest in whether greater technical specification would improve the precision of a paper's framing or claims.

\paragraph{Underspecification}
A paper is underspecified if any empirical strand is coded \emph{underspecified}: the paper gives too little detail to place the system it studies within our typology (\secref{sec:typology}).

\paragraph{Mischaracterisation}
A paper is mischaracterised if at least one motivation strand invokes a system type that differs \emph{consequentially} from the one studied -- where the mismatch, not merely the wording, matters for the claim being motivated.

\paragraph{Overgeneralisation}
A paper is overgeneralised if at least one conclusion strand reaches beyond the system type its empirical strands support, and the gap matters for the claim's validity or policy relevance. This is the costliest failure for cumulative science: later work may build on claims that do not hold across technical contexts \citep{schroederBigDataCumulation2020}.

\paragraph{LLM-Assisted Extraction} We use LLM-assisted extraction to support the initial extraction of candidate strands \citep{daiLLMintheloopLeveragingLarge2023,nguyen-trungChatGPTThematicAnalysis2025}. The LLM is used to impose a consistent preliminary structure for each paper; all coding decisions are made exclusively by human coders. Each paper is first converted into a full-text markdown representation and provided to the LLM together with the full coding prompt reproduced in \ref{app:coding}. We use Gemini 3.1 Flash-Lite Preview \citep{geminiteamGeminiFamilyHighly2025} to produce a structured extraction for each paper. After receiving the LLM output, the assigned human coder reads the full paper and revises, adds, merges, or removes strands as required. Coders independently make all judgements regarding system classification, public-value dimensions, mischaracterisation, and overgeneralisation, and do not receive LLM-generated suggestions, such that reported rates depend on human judgement alone.

The appendix codebook (\ref{app:full-coding}) is the prompt used for LLM-assisted extraction, reproduced verbatim. It defines both the preliminary extraction task given to the LLM and the annotation guidance used by human coders. The codebook specifies the typology labels, the public-value dimensions \citep[following][]{stalenhoefEenDialoogVoor2024}, the definition of each strand type, and the decision rules for identifying consequential mischaracterisation and overgeneralisation.

\paragraph{Scheme Validation and Refinement} The coding scheme was piloted on a subset of 10 papers coded by all authors, after which the codebook was refined to improve conceptual clarity and consistency \citep{mayringQualitativeContentAnalysis2015,schreierQualitativeContentAnalysis2012}. The remaining papers were randomly assigned to authors for independent coding. Ambiguous cases were recorded during coding and, after reliability assessment, discussed among the authors and resolved by consensus. These consensus decisions form the final dataset used for the analysis below.

\paragraph{Adjudication}
The judgements driving our paper-level outcomes are validated through codebook-grounded adjudication \citep{krippendorffContentAnalysisIntroduction2019}.\footnote{A double-coded subset large enough to estimate inter-coder agreement with usable precision was not feasible given corpus size and per-paper coding cost; a power analysis is provided in the repository.} For each paper, a second author re-assesses every strand whose value sets a paper-level flag -- motivation strands flagged as mischaracterised, conclusion strands as overgeneralised, and the empirical strands of any underspecified paper -- against the codebook and the paper text, retaining a flag only where its documented decision rule is met and removing it otherwise. Residual disagreements are settled by a third author \citep{oconnorIntercoderReliabilityQualitative2020}. 

We design adjudication conservatively: second coders can only remove imprecision flags set by the first coder, not add novel ones. Overturned flags lower the reported rate, while uncounted misses can only raise the true rate \citep{beggAssessmentDiagnosticTests1983}. The reported figures are therefore a conservative estimate of the prevalence of imprecision in the corpus. However, our design trades off against reviewer blinding: because adjudication is triggered by a flag, the second coder knows an error was proposed.

Of the 148 strands flagged by the primary coder, 128 were retained on adjudication, and 20 (14\%) were overturned. This non-trivial but modest rate is consistent with adjudication working as a genuine refinement.


\section{Findings} \label{sec:findings}

\begin{figure}
    \centering
    \includegraphics[width=1.0\linewidth]{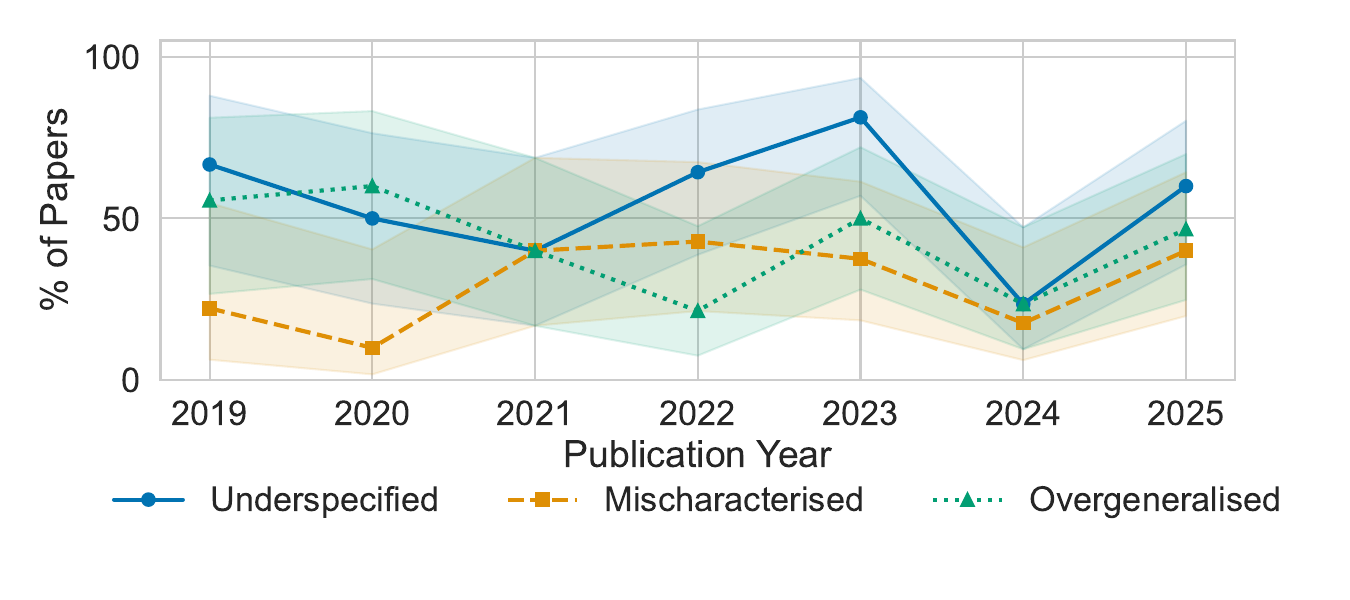}
    \caption{\textbf{Trends in rates}. We find no significant changes in any specification category over time.}
    \label{fig:error-year}
\end{figure}

We find significant imprecision across all three analysed categories. The results below present the overall rates and their relation to public values and typology dimensions. We find no changes in rate over time (Fig. \ref{fig:error-year}).
Summary statistics and figure-generation scripts are available in the project 
repository.\footnote{\url{https://anonymous.4open.science/r/AITypology4PA-4FC6/}}

\subsection{Underspecification} \label{sec:findings:underspecify}
\begin{figure}
    \centering
    \includegraphics[width=1.0\linewidth]{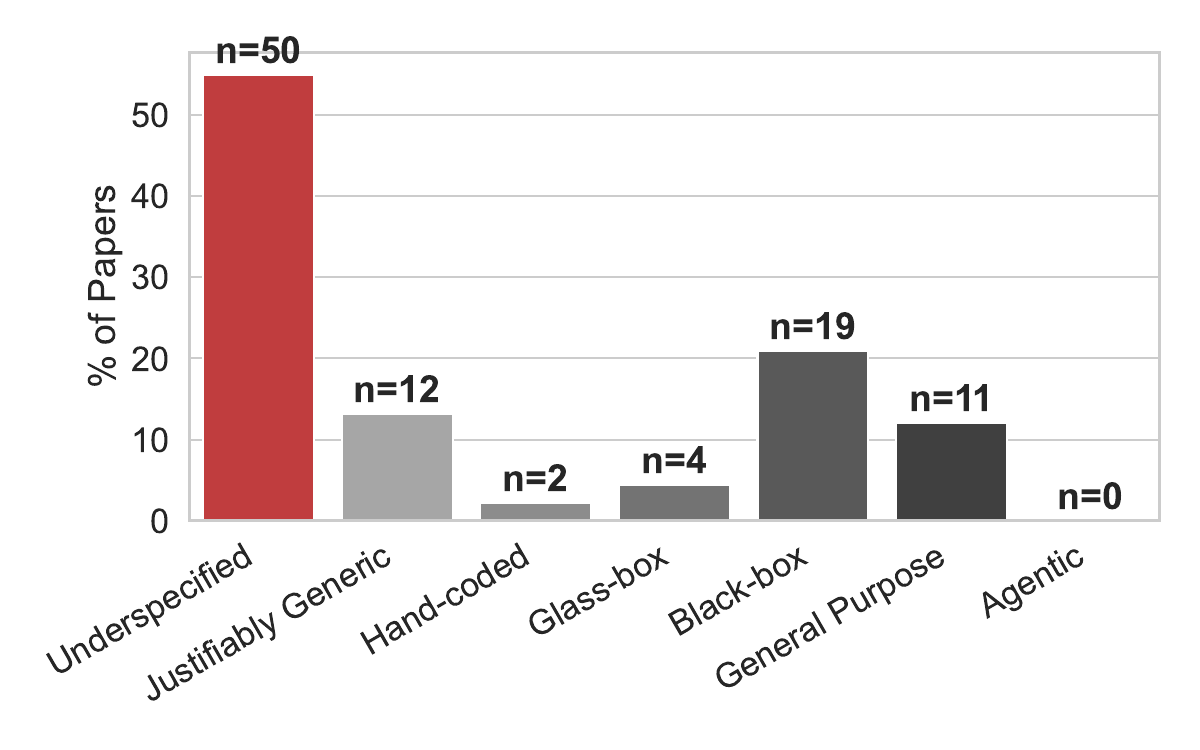}
    \caption{\textbf{Underspecification.} 55\% of papers provide insufficient information to determine which system is empirically studied. The most commonly studied system is black-box systems, with agentic systems completely unstudied.}
    \label{fig:underspecification}
\end{figure}

Of 91 coded papers, 50 ($55\%$) are underspecified: across all empirical references to the studied system, there is insufficient information to classify it with certainty. Fig. \ref{fig:underspecification} shows the number of analysed papers empirically studying each type of system in our typology.

Among fully specified papers, black-box systems are the most commonly studied category (N=19). In contrast, general-purpose systems are relatively understudied. Only 11 papers explicitly analyse general-purpose systems empirically, despite their growing prominence \citep{straubArtificialIntelligenceGovernment2023}.
No papers are classified as studying agentic systems in our corpus. Papers mentioning `agents' primarily engage with these systems at a conceptual level \citep[e.g., ``cognitive robots'' in][]{wirtzArtificialIntelligencePublic2019} or in relation to physical automation \citep[e.g., drones in][]{straubAIBureaucraticProductivity2024}, rather than contemporary LLM-based agents \citep{ilvesAgenticStateRethinking2025}. 
However, as our citation-weighted sampling structurally disadvantages recent work (\secref{sec:limitations}), some of this absence could reflect citation lag, as discussed in \secref{sec:discussion:agent}.

\subsection{Mischaracterisation}
\begin{figure}
    \centering
    \includegraphics[width=1.0\linewidth]{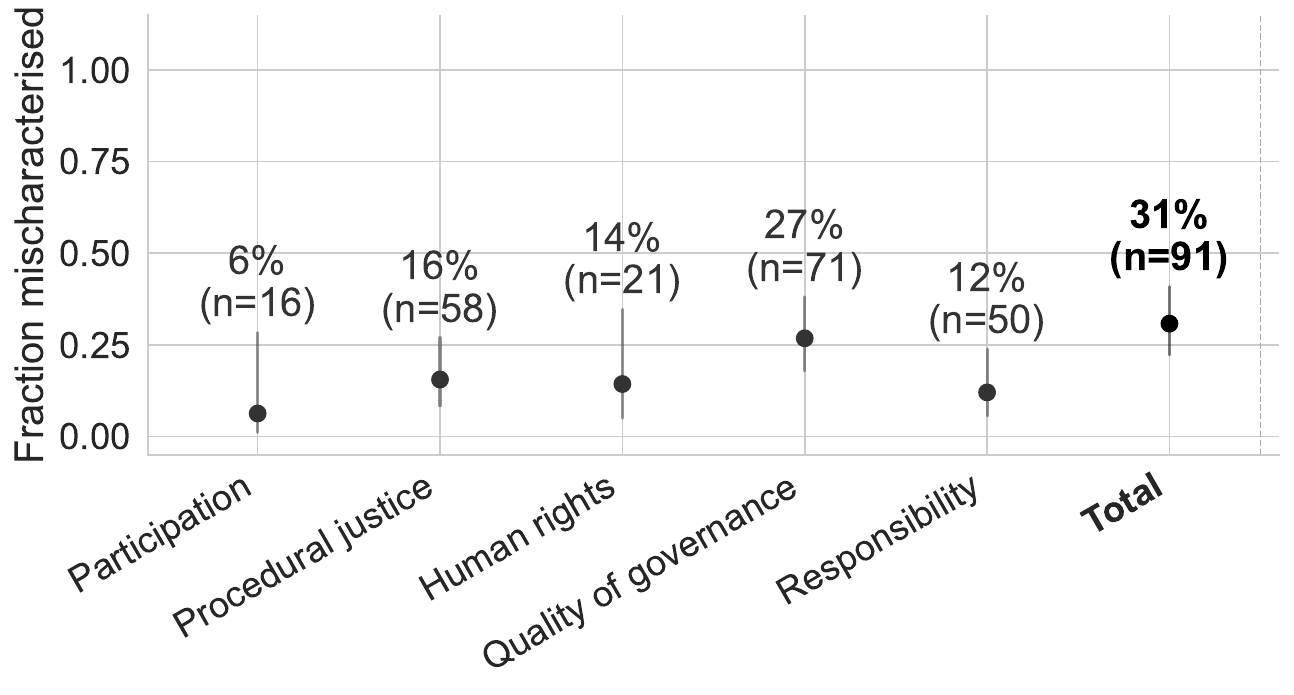}
    \caption{\textbf{Mischaracterisation.} Proportion of papers that have mismatches between systems mentioned in the \emph{motivation} and the systems \emph{empirically} studied. In total, 31\% of papers have mischaracterised strands. Error-bars are 95\% \citet{wilsonProbableInferenceLaw1927} scores.}
    \label{fig:mischaracterisation}
\end{figure}

31\% of coded papers mischaracterise AI systems: they exhibit at least one consequential mismatch between motivating and empirically analysed systems.

Fig. \ref{fig:mischaracterisation} shows the proportion of papers that make at least one mischaracterised claim within each governance dimension. We see statistically similar rates across value dimensions.



\subsection{Overgeneralisation}

\begin{figure}
    \centering
    \includegraphics[width=\linewidth]{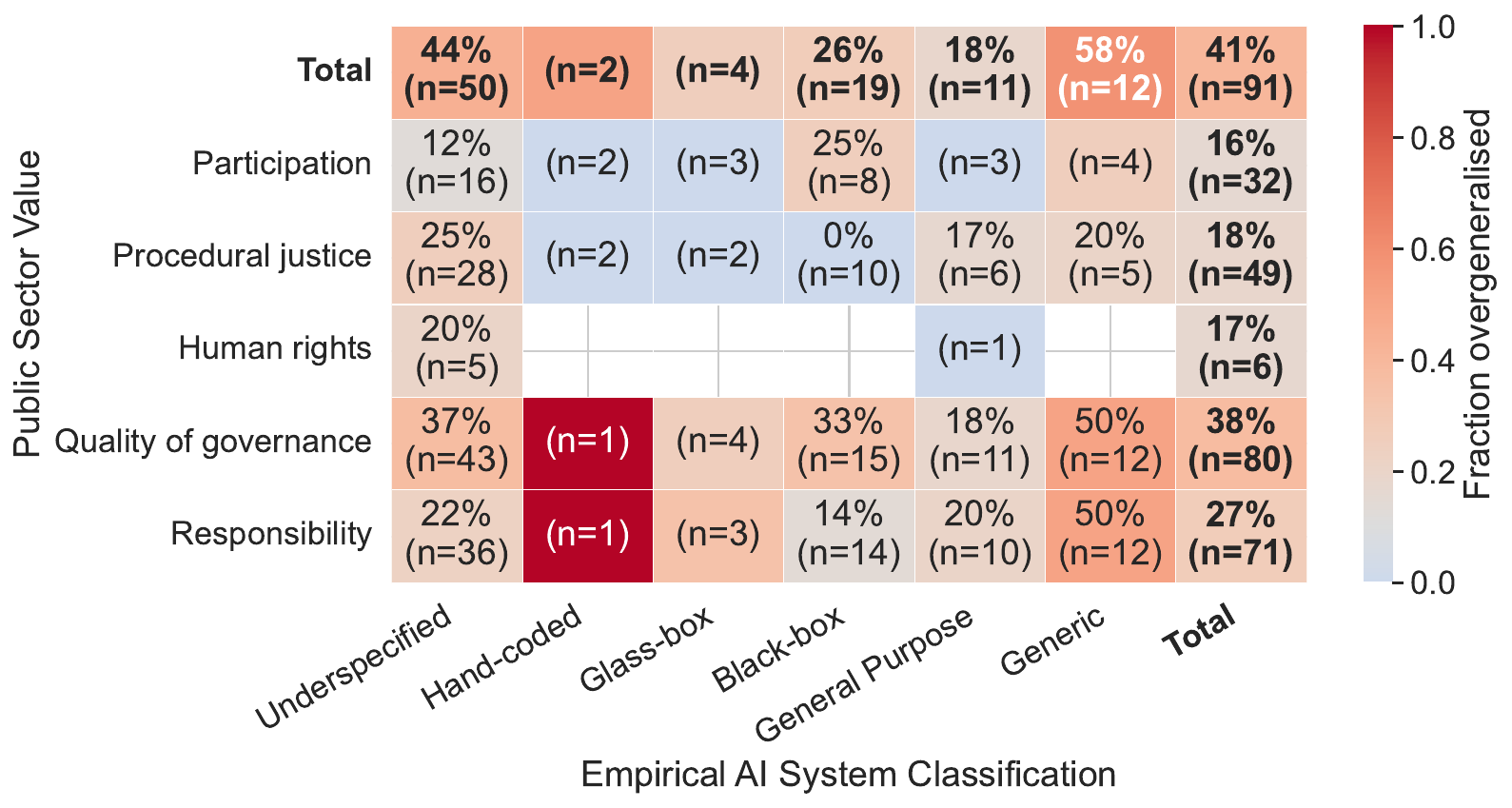}
    \caption{\textbf{Overgeneralisation.} Heatmap between overgeneralisation for system type (X-axis) and public value (Y-axis). Outer cells indicate marginals. In total, 41\% of papers overgeneralise.}
    \label{fig:overgeneralisation}
\end{figure}

41\% of coded papers make at least one claim which is more general than their empirics justify. Fig. \ref{fig:overgeneralisation} maps instances of overgeneralisation across typology dimensions and public values. We find significant rates of overgeneralisation in every cell with enough data to make statistical claims.

Generally, papers with underspecified systems (column 1), or that address ``AI'' generically (column 6), are more likely to make overgeneralised claims. The only exception is black-box systems (middle column), which also has a high prevalence. We discuss this further in \secref{sec:generic}.

Claims about the quality of governance are most likely to be overgeneralised. This category covers practical claims about implementation, such as organisational factors in AI use, or the tasks for which AI systems are used. These vary more frequently across technically different systems than the more fundamental and conceptual claims in other public-value categories.


\section{Discussion} \label{sec:discussion}

Our analysis indicates that public administration and digital government research about ``AI'' often overlooks technical distinctions that matter for governance. 
Sorting studied systems into a technical typology of just five categories suggests remarkable potential for more precision.
As developed in our theory (\secref{sec:theory}), such imprecision should be avoided because it harms the field's development of cumulative knowledge. 

We therefore make two contributions with the aim of improving the technical precision of future work on AI in public administration. First, in \secref{sec:pitfalls}, we highlight three common types of pitfall we find in our analysis -- both to illustrate practically how these harm knowledge development, and to help researchers avoid them in the future. Second, in \secref{sec:recommendations} we give practical recommendations for future research on AI in the public sector, which we believe greatly help technical precision -- without requiring researchers to have either deep technical knowledge or closer access to studied systems.

\subsection{Patterns of Imprecisions}
\label{sec:pitfalls}

Across the analysed papers, we find three prominent patterns of imprecision. These include confusion introduced by the use of generic terms (\secref{sec:generic}), overreliance on research about black-box systems (\secref{sec:too-black-box}), and a failure to ``future-proof'' claims, evidenced by their inapplicability to agentic systems (\secref{sec:discussion:agent}). 

\subsubsection{Generic Terms}
\label{sec:generic}

The single biggest driver of technical imprecision we find is the indiscriminate use of broad, generic, or ambiguous terms, such as ``AI", ``machine learning", ``algorithmic decision-making" (ADM), or ``chatbot". Three types of issues result.

First, most broad terms can refer to systems across the typology, such that they invite overgeneralisation -- in other words, authors use generic language but refer to specific systems. For example, \cite{davidPublicPerceptionsResponsible2025} attribute to AI a set of ``distinguishing features''  --  adaptive capacity, management of complex tasks, automation of decisions  --  without specifying which systems have them; the claim cannot be assessed because the referent is left open. \cite{andrewsPublicAdministrationPublic2019} similarly conflates ``algorithms'', which conventionally span hand-coded and learned systems, with ``machine learning'', collapsing a threshold across which transparency and accountability differ sharply (see \ref{sec:typology}). 

Second, many of these terms have imprecise or contested definitions in themselves. Most notably, as discussed above, ``AI'' is taken by some authors to include complex, but hand-coded rule-based systems, such as robotic process automation (RPA), while others take it as synonymous with ``machine learning'' -- covering only the second tier in our typology onwards. 
Combined with underspecification, such ambiguity can even cast doubt on whether studied systems are ``AI'' at all, and therefore on the AI-specificity of derived claims. 
Surveys of public-sector documents highlight this same issue in registers of federal AI applications \citep{khanHowPromoteAI2024} and AI policy initiatives \citet{noordtPolicyInitiativesArtificial2025}.
Where authors do not resolve these ambiguities, it is unclear what system types they draw from and map to.

Finally, the conception of some terms has advanced as technology has progressed. Take the term ``chatbot'': although the conversational user interface has remained similar, in the past decade chatbots have evolved from hand-coded ``conversation tree'' systems to generally capable, general-purpose-powered agents \citep{adamopoulouChatbotsHistoryTechnology2020,ilvesAgenticStateRethinking2025}. Reducing these vastly different systems to their interface is imprecise.  For example, \citet{aokiExperimentalStudyPublic2020} studies chatbots they label ``narrow AI'' without establishing whether the chatbots follow hand-coded conversation trees or use black-box NLP intent recognition. \cite{juCitizenPreferencesGovernment2023} note the higher fluency of GPT-like systems but design guidelines on assumptions that predate the externalisation these systems presuppose. It appears plausible that this imprecision is driven by the term ``chatbot'' being established even as the affordances of the underlying technology have changed drastically.

Where these terms are defined and scoped clearly, their use can, of course, be appropriate: for example, discussion of the ``intransparency of AI'' may hold across all systems learned from data \citep{bullockArtificialIntelligenceBureaucratic2020, lazarLegitimacyAuthorityDemocratic2024}. It may even be required to use such terms, to reflect analysis of their use or perception: vignette experiments, for example, may reasonably describe a system as ``AI-based software'' to test what participants infer. But derived claims can still overgeneralise: \cite{geskArtificialIntelligencePublic2022} state that technical classifications ``are therefore not elaborated here'', despite motivating the study with the opacity and undocumented rules of black-box systems  --  affordances its generic stimulus never instantiates.


\subsubsection{Overextrapolation from Black-Box Systems}
\label{sec:too-black-box}

A second common pitfall is overextrapolation of conclusions that were drawn based on the study of black-box systems. Black-box systems are prominent: they are the most studied category and the empirical basis for many conclusions about other layers (Figs.~\ref{fig:underspecification},~\ref{fig:overgeneralisation}); where papers underspecify the studied system, we most frequently speculate that it is black-box. However, the affordance profile of black-box systems is narrow: they are usually trained separately by each organisation on their own data, purpose-bound, and they produce numeric or binary outputs, such as risk scores, likelihood estimates, or yes/no decisions.

Imprecisions frequently result from overextending claims made about black-box systems. For example, \cite{wangProvinceOriginDecisionmaking2025} draw general conclusions about ``algorithmic'' decision-making from a study whose effects on participation plausibly depend on the level of transparency, abandoning the rule-driven/data-driven distinction the same authors drew in \cite{wangWhatTypeAlgorithm2023}. \cite{wirtzArtificialIntelligencePublic2019} present opacity, training-data bias, and autonomous learning as universal challenges of AI, even though their own application table includes rule-based systems to which these do not apply; their claims about implementation capacity and accountability hold cleanly only for the black-box layer. \cite{chenAdoptionImplementationArtificial2024} extend a functional typology \citep{makasiTypologyChatbotsPublic2022} developed before the proliferation of general-purpose models, but do not register the change in the skills required to audit and govern such systems \citep{mokanderAuditingLargeLanguage2024}.

These overextensions span most public value dimensions, but often share three patterns. First, claims on \textit{participation} and \textit{procedural justice} are often only valid for systems with the \textit{explainability} affordance of black-box systems. These produce a singular, quantitative output, and ``explainability'' is taken to mean an understanding of model internals that produce it, e.g. generated via explainable AI (XAI) techniques \citep{mowbrayExplainableAIXAI2023}. In contrast, LLMs may produce long text outputs -- which can contain testable explanations in themselves, and therefore be institutionally valid without any understanding of model internals \citep{schmitzMoralAgencyFramework2025}. 

Second, claims on \textit{quality of governance} are often over-indexed on the technical or organisational specifics of black-box models. For example, large volumes of high-quality data are often named as a requirement to ``train AI'', 
but externally procured GPAI systems do not require any internal training data. \cite{alon-barkatAlgorithmicDiscriminationPublic2025} find that in-house development raises perceived responsibility relative to outsourcing, but treat internalisation as a free choice  --  whereas general-purpose systems carry inherent externalisation pressures relevant for implementation capacity, so the finding may not hold where the model is developed elsewhere (\secref{sec:typology:general}). 

Last, claims on \textit{responsibility} from black-box models can underestimate the complexity of accountability allocation in modern AI supply chains \citep{brownAllocatingAccountabilityAI2023}. Black-box systems invite the assumption that data and model training are both internal to the organisation. Further, there is a difference in the type of AI system outputs citizens and officials interact with: an LLM-generated text explanation, for example, may be more persuasive to a decision-maker than a single numeric score \citep{salviConversationalPersuasivenessGPT42025}, calling into question conclusions about, e.g., automation bias \citep{alon-barkatHumanAIInteractions2023}. For example, \cite{keppelerHowEnsemblingAI2025} study human--AI ensembles using a black-box tool but generalise their conclusions to ``AI advice'' in general.

The overreliance on black-box systems likely has historical drivers. Much of the fundamental literature on ``AI in government'' was published between 2019 and 2022 \citep{aarabIntegratingAIPublic2025}, when such systems formed the frontier of AI capabilities \citep{brownLanguageModelsAre2020}. Indeed, a black-box quantitative risk scoring model likely caused the canonically referenced Dutch benefit scandal, which spurred an explosion of work in the field  \citep{peetersAdministrativeExclusionInfrastructurelevel2023}. The 2022 ``general-purpose shift'' driven by the introduction of ChatGPT then introduced a new class of system with drastically different affordances (\secref{sec:typology}) and regulatory and societal implications \citep{wangDistinguishingTaskspecificGeneralpurpose2026} -- shortly after the canon developed.

\subsubsection{Inapplicability to Agentic Systems (``Future-proofing'')} \label{sec:discussion:agent}

A third form of imprecision we find is failure to address agentic systems, the newest layer of the typology. This takes two forms: some claims made in work published before agentic systems proliferated do not translate to them, and the field empirically so far does not study their deployment.

The shift from general-purpose to agentic systems affects affordances across all public values (\secref{sec:typology:agentic}), but most consequentially \textit{responsibility}, because of the implications for human oversight. Moving from reactive input-output mapping to proactive action over time (\secref{sec:typology:agentic}) moves oversight from the ex-post evaluation of discrete outputs to the runtime monitoring of ongoing processes \citep{schmitzOversightStructuresAgentic2025,chanVisibilityAIAgents2024}. Accountability must be allocated for extended courses of action, rather than in discrete moments of decision, diffusing discretionary power away from identifiable actors \citep{chanHarmsIncreasinglyAgentic2023,schmitzMoralAgencyFramework2025}. Findings about the accountability of general-purpose chatbots -- where a human can review each output -- do not transfer to agentic systems that act across system boundaries without per-step review, since the oversight point has moved. 

We flag imprecisions in many papers because they make general claims about ``AI'' that are invalidated by this affordance boundary. For example, as \citet{busuiocAccountableArtificialIntelligence2021} highlights, whether technical transparency solves accountability questions is a question of bureaucratic and process design. That interventions such as XAI improve perceived accountability for single-point decisions, therefore, does not express anything about their impact on the accountability of multi-turn agent actions.

Further, across the reviewed papers, we find no study of agentic systems themselves (see Fig. \ref{fig:underspecification}). 
Given the recency of these systems, this is unsurprising -- technical research on agents is accumulating, but little of it speaks to public administration \citep{rystromAgentBenchmarksFail2026}.



Beyond the specifics of agentic systems, this failure mode highlights how technical precision also contributes to making claims ``future-proof'': as AI systems change and improve, claims about ``AI'' are more likely to age poorly than those with clear system types. For example, \cite{wangWhatTypeAlgorithm2023} experimentally compare rule-driven (hand-coded) and data-driven (black-box) decision-making, and \cite{keppelerNoThanksDear2024} likewise grounds its study of disclosure effects in black-box systems. Both of these remain valuable contributions, and it is clear how their insights map to agents.


\subsection{Recommendations for Public Administration Research} \label{sec:recommendations}

Our work demonstrates that PA researchers should strive to improve the durability and generalisability of their findings by being more technically precise about AI. However, in so doing, they may encounter practical challenges: access to detailed information can be difficult, they may rely on surveys or interviews with non-experts, or they may lack the necessary technical background. We provide three sets of practical recommendations.

\begin{tcolorbox}[colback=blue!5!white, colframe=blue!30!white, fonttitle=\bfseries\small, title={Recommendations for Specifying AI Systems}, left=4pt, right=4pt, top=3pt, bottom=3pt, before upper={\small}]
\begin{enumerate}
    \item \textbf{Explicitly Specify AI System Types}
    \begin{itemize}
        \item Situate the system under study within a structured typology, such as the one presented here. Its design serves as a specification checklist: answering the four diagnostic questions in Fig.~\ref{fig:typology} places a system in exactly one class.
        \item Add as much technical detail as necessary to clarify the affordances of the system, e.g. the specific name of studied LLMs -- but no more.  
        \item Consider including a concrete diagram, system visualisation, or practical example of the system in use, helping readers quickly assess the system's affordances and scope.
    \end{itemize}
    \item \textbf{Use Proxy Indicators and Flag Uncertainty} 
    \begin{itemize}
        \item Where technical detail is unavailable, approximate the affordances of the system with proxy indicators, such as the data used to train the AI model or its precise type of inputs and outputs. 
        \item Explicitly highlight any remaining uncertainty about technical specifics, rather than generalising to ``AI''.
    \end{itemize}
    \item \textbf{Scope Relevance of Past Work and Conclusions}
    \begin{itemize}
        \item Before drawing on past work, attempt to determine the AI system studied in it, and judge whether its affordances allow meaningful translation. 
        \item When drawing conclusions, be explicit about what types of AI systems you expect your claims to generalise to.
    \end{itemize}
\end{enumerate}
\end{tcolorbox}

\paragraph{1. Explicitly specify AI system types}
Scholars should specify the type of AI system they study, such as by placing it in the typology we propose. This does not require exhaustive technical detail, just enough specificity for readers to understand the affordance profile of the system. In Fig.~\ref{fig:typology}, we provide four diagnostic questions. Answering these top-to-bottom maps an AI system to exactly one class. For typical PA cases, each of these is answerable from publicly available information about the system as deployed  --  without access to source code or model architecture.

Our typology as presented is a minimum bound on technical specificity (\secref{sec:theory}), but for some topics, more technical detail may be warranted. Many systems also combine layers  --  a black-box system embedded in a hand-coded decision system, say. While we discuss how the typology could be expanded below (\secref{sec:further-research}), individual authors may use a simple affordance-based litmus test to decide how much detail to include: \textit{would adding this detail distinguish between two systems with meaningfully different affordances}? 

For example, different LLMs perform differently on public-sector tasks \citep{rystromAgentBenchmarksFail2026}. Authors studying an LLM-based chatbot should therefore err towards naming the model used to clarify its affordances (specific to a version, e.g. ``Gemini 3.1 Flash-Lite Preview'', which we use above), rather than referring to ``an LLM''. 

\paragraph{2. Use proxy indicators and flag uncertainty}
Where practical challenges prevent the above specification, authors should a) use proxy indicators to approximate affordance profiles, and b) highlight any uncertainty that remains. Proxy indicators about AI systems may be available even if the above technical detail is not. These may include:

\begin{itemize}
    \item The type of data used to train the AI model, and who trained it.
    \item The way the AI model is hosted and accessed by the organisation (e.g. on-premise vs. remotely).
    \item The model's or system's input and output types -- such as a single risk score or a free-text explanation.
    \item Information about the system's performance, such as its classification accuracy or benchmark results.
    \item If the AI model or system is a third-party product, its name or vendor.
\end{itemize}

As we theorise (\secref{sec:typology}) and demonstrate (\secref{sec:pitfalls}), each of these indicators readily provides affordance-relevant information, and should therefore not be written off as irrelevant or overly technical. 

Finally, should uncertainty remain, describing that uncertainty is more informative than an undifferentiated generalisation to ``AI''. Doing so conveys the maximal intended scope of claims, eases (or allows) retroactive specification, and ``future-proofs'' statements.

\paragraph{3. Scope relevance of past work and conclusions} 
Technical precision should not only be applied to the AI system at hand: researchers should apply similar precision both when drawing on past work on AI in PA, and when concluding beyond the systems studied.

To avoid mischaracterised motivation, researchers should attempt to typologise the AI systems which past work studies, and judge whether core claims translate. For example, a paper on algorithmic transparency studying black-box systems may provide valuable framing for a paper studying a general-purpose system, but the exact transparency techniques employed may not translate. As our methodology shows (\secref{sec:analysis}), such analysis is possible retroactively in many cases.

Similarly, scholars should specify for which types of AI systems they expect their conclusions to hold. If scoped well, conclusions can evidently be more general than the single case or system studied. Public administration is deeply familiar with phrasing such scope conditions: scholars are careful about whether findings depend on a particular institutional setting, administrative tradition, policy sector, or level of government. The same practice should be commonplace for technical reach. To ``future-proof'' claims, a practical solution may be to scope them to ``currently available'' AI systems.

\subsection{Further Research}
\label{sec:further-research}

Our typology serves two purposes: it exemplifies in general that technical precision about AI beyond the current standard is necessary, and it enables such precision for \textit{current} systems. This focus suggests two promising strands for future research.

First, it may be fruitful to detail out the typology we introduce -- both ``horizontally'' by adding more nuanced public-value dimensions, and ``vertically'' by distinguishing more granularly between system types. For example, agentic systems have ``degrees of agenticness'' \citep{kasirzadehCharacterizingAIAgents2025} and vary in their autonomy, goal-directedness, and impact. These degrees may affect the public-sector affordances that different agentic systems have.

Second, as AI systems evolve, research on their public-value implications should keep pace. Newer systems may have novel affordance profiles compared to current ones. For example, three potentially consequential developments in AI research are the \textit{increasing agenticness} of AI systems discussed above, \textit{continual learning} methods -- which produce AI systems whose internal structure constantly updates, rather than being static after training \citep{yuRecentAdvancesMultimodal2026}, and \textit{embodiment}, the integration of general-purpose systems with physical hardware \citep{firooziFoundationModelsRobotics2025}. 
Each of these advances, and others that may emerge, could produce systems with novel affordance profiles, and PA research should analyse how these match or differ from past ones.

\subsection{Limitations}
\label{sec:limitations}

Beyond possible extensions in future research, we highlight three possible limitations of our work.

\paragraph{Case Selection} Our analysis draws on a specific sample -- the most highly cited papers shaping public administration and digital government scholarship on AI (2019--2025) -- which may not represent the field as a whole. Our sample may exhibit different patterns than one composed of less-cited or more applied work. 
Specifically, citation-weighted sampling may over-represent conceptual and review work relative to applied case studies (Table~\ref{tab:appendix_corpus_summary}). It also structurally disadvantages recent work, which may partly explain the scarcity of papers on general-purpose and agentic systems. 

\paragraph{Coding} Because our flagging is conservative -- every positive is adjudicated by a second coder -- the reported mischaracterisation and overgeneralisation prevalences are also conservative. 
Further, coding errors remain possible despite our measures to prevent them: we report adjudication rates and a power analysis, and reach no unresolved disagreement about codes in adjudication. 


\paragraph{Detail and Currency} As discussed above (\secref{sec:further-research}), there are still unexplored implications of our typology, and it will require updating as novel AI systems are introduced. We are explicit about these bounds and suggest both directions for future work. 



\section{Conclusion} \label{sec:conclusion}

The expansion of AI in public administration has spurred a robust and valuable body of research. As our structural review demonstrates, this existing literature provides an essential foundation for understanding how algorithmic systems interact with core public values such as democratic participation, procedural justice, and governing capability. However, the conceptual tools used to classify these systems must keep pace with their technological architectures without getting swept away by a torrent of technical distinctions. 


But stronger technical specification of AI system types is a worthwhile investment. Retaining the umbrella term ``AI'' without technical clarification produces underspecification, internal inconsistency, and overgeneralisation that weaken otherwise sound findings. 

Avoiding these methodological pitfalls does not require public administration scholars to adopt highly granular engineering taxonomies. It only requires anchoring our definitions to affordance thresholds -- the points at which a technical shift fundamentally alters what governance actors can or cannot do. By applying just a slight increase in specificity, researchers can significantly extend the transferability and applicability of their claims, ensuring that insights drawn from one context are reliably mapped to the right systems in the future.


\section*{Declaration of generative AI and AI-assisted technologies in the manuscript preparation process}
Large language models are a central part of the methodology as described in \secref{sec:analysis}. Specifically, we use Gemini 3.1 Flash-Lite to extract structured information as part of our qualitative coding pipeline. All judgments and assessments were made solely by the authors, with no LLM-generated suggestions. All author judgements and LLM-extracted strands are available in the project repository.

Furthermore, Claude Code was used to assist in creating the plots and figures. All code was reviewed and validated by the authors. ChatGPT and Claude were used for light copy-editing. The authors take full responsibility for all content and materials.

\newpage
\bibliographystyle{elsarticle-harv} 
\bibliography{zotero2.bib}

\appendix

\usetikzlibrary{arrows.meta, positioning, shapes.geometric}
\section{Data, coding, and corpus summary} \label{app:data}
This appendix details corpus construction and coding. Figure~\ref{fig:prisma_flow} summarises the screening process and Table~\ref{tab:appendix_coding_dimensions} summarises the coding dimensions. The corpus was assembled from all venues listed in the study configuration (Table~\ref{tab:venues}), using OpenAlex as the retrieval source, a title- and abstract-based keyword filter, and a citation-based annual sampling rule. The OpenAlex query string, venue list, and corpus metadata are provided as machine-readable files at \url{https://anonymous.4open.science/r/AITypology4PA-4FC6/}.

\begin{table*}[ht]
  \centering
  \caption{Journals included in the literature review}
  \label{tab:venues}
  \begin{tabular}{ll}
    \toprule
    \textbf{Stream} & \textbf{Journal} \\
    \midrule
  \multirow{12}{*}{Public Administration} & \href{https://portal.issn.org/resource/ISSN/0033-3352}{Public Administration Review} \\
   & \href{https://portal.issn.org/resource/ISSN/1477-9803}{Journal of Public Administration Research and Theory} \\
   & \href{https://portal.issn.org/resource/ISSN/0033-3298}{Public Administration} \\
   & \href{https://portal.issn.org/resource/ISSN/0952-1895}{Governance} \\
   & \href{https://portal.issn.org/resource/ISSN/1471-9037}{Public Management Review} \\
   & \href{https://portal.issn.org/resource/ISSN/2515-4303}{Perspectives on Public Management and Governance} \\
   & \href{https://portal.issn.org/resource/ISSN/0275-0740}{American Review of Public Administration} \\
   & \href{https://portal.issn.org/resource/ISSN/0095-3997}{Administration \& Society} \\
   & \href{https://portal.issn.org/resource/ISSN/0190-0692}{International Journal of Public Administration} \\
   & \href{https://portal.issn.org/resource/ISSN/1530-9576}{Public Performance \& Management Review} \\
   & \href{https://portal.issn.org/resource/ISSN/0276-8739}{Journal of Policy Analysis and Management} \\
   & \href{https://portal.issn.org/resource/ISSN/0020-8523}{International Review of Administrative Sciences} \\
  \midrule
  \multirow{4}{*}{Digital Government} & \href{https://portal.issn.org/resource/ISSN/0740-624X}{Government Information Quarterly} \\
   & \href{https://portal.issn.org/resource/ISSN/1570-1255}{Information Polity} \\
   & \href{https://portal.issn.org/resource/ISSN/1548-3886}{International Journal of Electronic Government Research} \\
   & \href{https://portal.issn.org/resource/ISSN/2639-0175}{Digital Government: Research and Practice} \\
  \bottomrule
  \end{tabular}
\end{table*}

\subsection{Corpus screening}
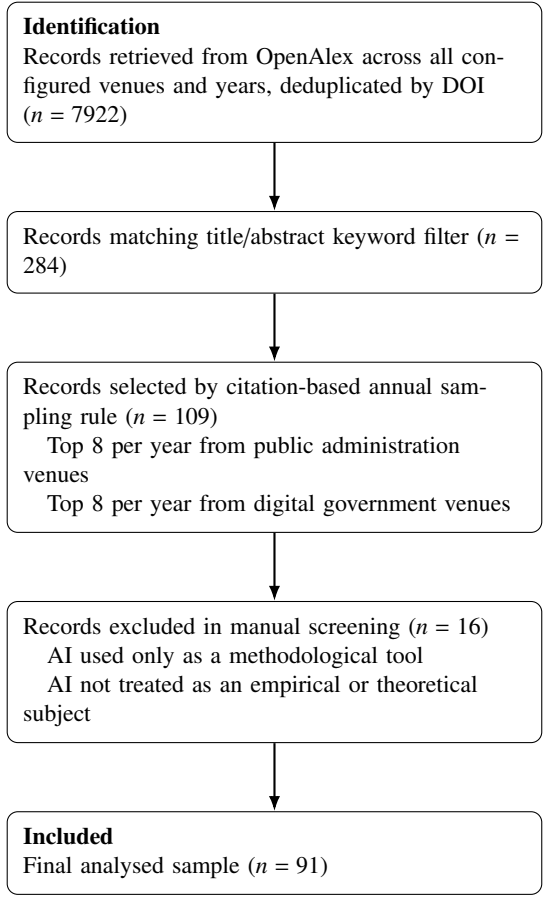
\begin{figure}[t]
\centering
\begin{tikzpicture}[
  node distance=0.9cm,
  every node/.style={font=\small},
  box/.style={
    rectangle,
    rounded corners,
    draw,
    align=left,
    text width=0.75\columnwidth, 
    inner sep=6pt
  },
  arrow/.style={-{Latex[length=2mm]}, thick}
]

\node[box] (id) {%
\textbf{Identification}\\
Records retrieved from OpenAlex across all configured venues and years, deduplicated by DOI ($n=7922$)
};

\node[box, below=of id] (match) {%
Records matching title/abstract keyword filter ($n=284$)
};

\node[box, below=of match] (sample) {%
Records selected by citation-based annual sampling rule ($n=109$)\\
\hspace*{1em}Top 8 per year from public administration venues\\
\hspace*{1em}Top 8 per year from digital government venues
};

\node[box, below=of sample] (exclude) {%
Records excluded in manual screening ($n=16$)\\
\hspace*{1em}AI used only as a methodological tool\\
\hspace*{1em}AI not treated as an empirical or theoretical subject
};

\node[box, below=of exclude] (final) {%
\textbf{Included}\\
Final analysed sample ($n=91$)
};

\draw[arrow] (id) -- (match);
\draw[arrow] (match) -- (sample);
\draw[arrow] (sample) -- (exclude);
\draw[arrow] (exclude) -- (final);

\end{tikzpicture}
\caption{Corpus construction and screening procedure. Articles were retrieved from all configured venues using OpenAlex, deduplicated, filtered using title- and abstract-based keyword matching, and then sampled using a citation-based rule selecting the eight most-cited papers per year separately for public administration and digital government venues. The resulting set was manually screened to exclude papers that used AI only as a methodological tool or did not treat AI as an empirical or theoretical subject.}
\label{fig:prisma_flow}
\end{figure}

The corpus was assembled in four stages. First, we retrieved all articles published between 2018 and 2025 from the configured public administration and digital government venues listed in \texttt{venues.yaml}. Second, retrieved records were deduplicated by DOI where available. Third, we applied a keyword filter to titles and abstracts reconstructed from the OpenAlex abstract inverted index. Fourth, from the matched set we selected the eight most-cited papers per year separately for public administration venues and digital government venues, to balance the corpus across the two journal streams. The resulting set was then manually screened to exclude papers that used AI only as a methodological tool or did not treat AI as an empirical or theoretical subject.

Figure~\ref{fig:prisma_flow} presents this process in PRISMA-like form \citep{pagePRISMA2020Statement2021}. A total of 7{,}922 records were retrieved from OpenAlex and deduplicated by DOI. Of these, 325 matched the keyword filter. Applying the citation-based annual sampling rule yielded 122 papers, from which 21 were excluded during manual screening, leaving a final analysed sample of 101 papers.

\begin{table*}[t]
\centering
\caption{Descriptive summary of the analysed corpus}
\label{tab:appendix_corpus_summary}
\begin{tabular}{p{11cm}r}
\toprule
\textbf{Panel A. Final sample composition} & \textbf{N} \\
\midrule
Final analysed papers & \textit{91} \\
Public administration venues & \textit{42} \\
Digital government venues & \textit{49} \\
\addlinespace[0.5em]
Published in 2018 & \textit{0} \\
Published in 2019 & \textit{9} \\
Published in 2020 & \textit{10} \\
Published in 2021 & \textit{10} \\
Published in 2022 & \textit{14} \\
Published in 2023 & \textit{16} \\
Published in 2024 & \textit{17} \\
Published in 2025 & \textit{15} \\
\midrule
\textbf{Panel B. Empirical paper-level system classifications} & \textbf{N} \\
\midrule
Hand-coded & \textit{2} \\
Glass-box & \textit{4} \\
Black-box & \textit{19} \\
General-purpose & \textit{11} \\
Agentic & \textit{0} \\
Underspecified & \textit{50} \\
\midrule
\textbf{Panel C. Paper-level outcome flags} & \textbf{N} \\
\midrule
Underspecified & \textit{50} \\
Mischaracterised & \textit{28} \\
Overgeneralised & \textit{37} \\
\midrule
\textbf{Panel D. Empirics type} & \textbf{N} \\
\midrule
Case study & \textit{22} \\
Survey & \textit{10} \\
Vignette experiment & \textit{21} \\
Experiment & \textit{4} \\
Systematic literature review & \textit{23} \\
Conceptual framework & \textit{10} \\
Other & \textit{1} \\
\bottomrule
\end{tabular}
\end{table*}

\subsection{Coding procedure and variables} \label{app:coding}

We conducted structured manual coding of all papers in the final sample. The coding scheme was jointly piloted by all three authors on 10 papers and refined iteratively before full coding began. Candidate quotations were surfaced using an LLM (see repository), after which one author read each paper in full and coded all relevant references using the final scheme. Multiple rows were created when a quotation referenced multiple systems, and multiple PA-relevance labels were allowed where applicable.

Each extracted reference was coded along three dimensions: \emph{AI system classification}, \emph{role in paper}, and \emph{PA relevance}. AI system classification used the typology described in the main text; role in paper distinguished Motivation, Empirical, and Conclusion; and PA relevance used the second-level dimensions of the good digital governance framework. Coders also recorded a brief justification for each code. Table~\ref{tab:appendix_coding_dimensions} summarises these dimensions.

\begin{table}[t]
\centering
\caption{Summary of coding dimensions}
\label{tab:appendix_coding_dimensions}
\begin{tabular}{p{2.6cm}p{4.2cm}p{1.2cm}}
\toprule
\textbf{Dimension} & \textbf{Values} & \textbf{Unit} \\
\midrule
AI system classification &
Hand-coded; Glass-box; Black-box; General-purpose; Agentic; Justifiably Generic; Underspecified &
Paper \\
\addlinespace
Role in paper &
Motivation; Empirical; Conclusion &
Strand \\
\addlinespace
PA relevance &
Participation; Procedural justice; Human rights; Quality of governance; Responsibility; None &
Strand \\
\bottomrule
\end{tabular}
\end{table}

The three paper-level outcomes reported in the main text were derived from these strand-level codings. A paper was classified as \emph{underspecified} if all of its empirical references were coded as Underspecified. A paper was classified as \emph{mischaracterised} if, within a given PA-relevance dimension, at least one motivation strand referred to a different system type from the one studied empirically, and as \emph{overgeneralised} if, within a dimension, at least one conclusion strand did so. Papers studying multiple empirical system types could contribute to multiple empirical categories. The paper-level results table and the analysis scripts implementing these rules are available in the repository linked above.

\section{Stability analysis of aggregate estimates} \label{app:saturation}

We assess sampling adequacy through a pre-specified stability analysis of the aggregate outcomes produced by our sampling rule. This is not a test of thematic saturation in the sense of determining whether additional papers would yield new concepts or codes. Instead, the analysis evaluates whether the paper's main aggregate findings are stable to the inclusion of additional papers within the same sampling frame. The procedure is therefore closer to stability-based sample-size assessment, where estimates are judged adequate once they remain within a specified tolerance corridor \citep{schonbrodtWhatSampleSize2013}, while also drawing on methodological work that operationalises saturation through explicit stopping rules and incremental sampling criteria \citep{francisWhatAdequateSample2010,guestSimpleMethodAssess2020}.
The bootstrap stability analysis is implemented in 
\texttt{saturation\_analysis.py}, available at 
\url{https://anonymous.4open.science/r/AITypology4PA-4FC6/}.

\begin{figure}
    \centering
    \includegraphics[width=\linewidth]{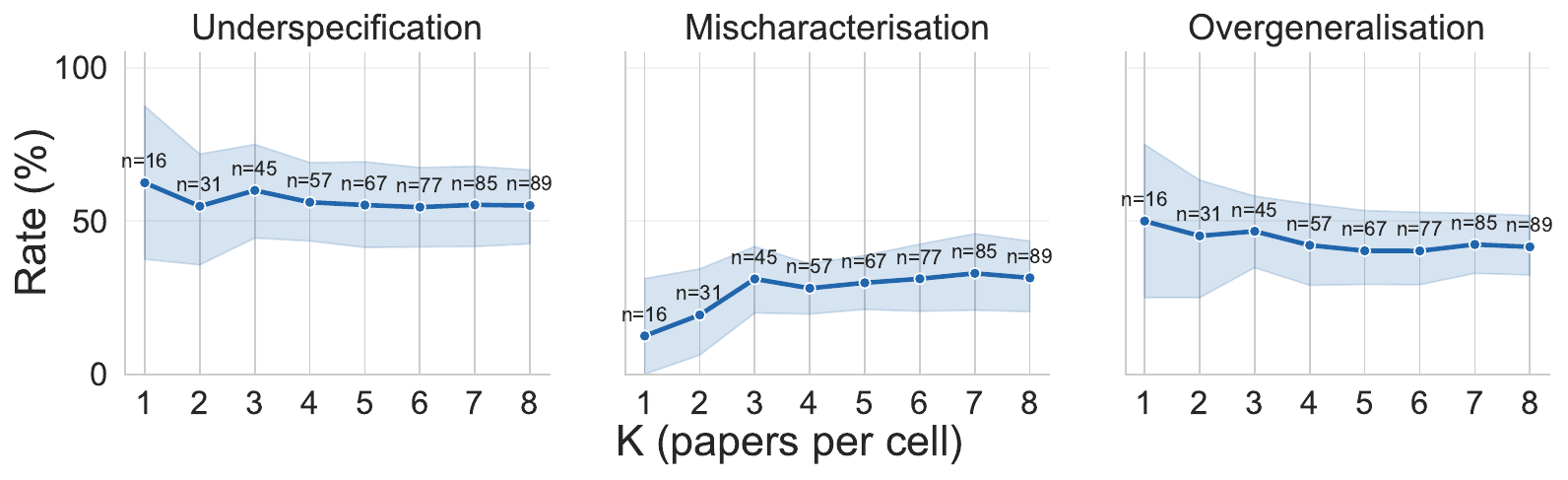}
    \caption{\textbf{Stability analysis.} Aggregate rates for our three main analytical constructs as we increase our sampling criteria. All constructs meet our stability criteria at $K=8$.}
    \label{fig:saturation}
\end{figure}

For each $K \in \{1,\dots,8\}$, we form a corpus by taking the top-$K$ papers from each stream-year combination, compute the three aggregate outcomes introduced in \secref{sec:operationalisation}---underspecification, mischaracterisation, and overgeneralisation---and apply two pre-specified criteria. First, \emph{local stability} requires the point estimates at $K=6,7,8$ to lie within 3 percentage points of each other. Second, \emph{flat trajectory slope} requires a linear fit over $K\in\{5,\dots,8\}$ to have a slope of at most 0.5 percentage points. These criteria operationalise the requirement that adding further papers within the sampling rule should not materially change the aggregate estimates. However, the exact values are somewhat arbitrary; the substantive evidence is the visual convergence as shown in Fig. \ref{fig:saturation}.

Uncertainty at each $K$ is calculated by block-bootstrapping stream-year combinations with replacement over 1{,}000 iterations to produce 95\% bands, following the general use of bootstrap resampling to quantify sampling variability around estimated quantities \citep{davisonBootstrapMethodsTheir1997}. Both criteria are met for all three outcomes at $K=8$---see Fig.~\ref{fig:saturation}. This provides an empirical bound on how much adding further papers within the same sampling rule would shift our aggregate findings, and supports treating $K=8$ as adequate for the substantive claims we make. The stability result is conditional on the sampling scope---top-cited PA and digital government venues, 2019--2025---and does not extend to claims about scholarship outside this scope.

\section{Codebook} \label{app:full-coding}
\begin{Verbatim}
You are analyzing a public administration research paper for **technical precision in its treatment of AI**. The core question throughout is: does the paper's use of "AI" as an undifferentiated category cause analytical problems — in its motivation, its empirical claims, or its conclusions?

---

### **YOUR TASK**

Produce a structured analysis covering three sections: **Empirics**, **Motivation**, and **Conclusions/Claims**. Do Empirics first, as your classification there anchors the judgements you make in the other two sections.

---

### **TAXONOMY OF AI SYSTEM TYPES**

Use this typology consistently throughout. When classifying systems, always ask: what is the most precise classification the paper's evidence actually supports? Default to a more conservative classification when in doubt.

### 1. Hand-coded

Traditional software systems where all rules and logic are explicitly written by humans. The system's behaviour is fully determined by its code and does not change based on data.

- **Key characteristics:** Explicit rules, stable behaviour, fully interpretable, no learned components  
- **Examples:** Rule-based benefit eligibility systems, tax calculation software, structured decision trees implemented as code  
- **Tip:** If the paper describes a system that applies fixed rules to determine outcomes (e.g. "the system decides who gets benefits based on income thresholds"), classify as Hand-coded.

---

### 2\. Glass-box systems

Machine learning systems that learn from data but produce interpretable, human-readable models. Experts can inspect and understand the model's decision logic.

- **Key characteristics:** Data-driven, interpretable output, simple learned rules  
- **Examples:** Logistic regression, linear regression, decision trees, PCA, TF-IDF scoring  
- **Tip:** If the paper describes a system as statistically trained but whose outputs can be understood or audited by experts, classify as glass-box.

---

### 3\. Black box systems

Machine learning systems that trade interpretability for performance. The internal logic of the model cannot be directly inspected, requiring post-hoc explanation methods.

- **Key characteristics:** High performance, opaque internal logic, operates on structured or unstructured data, post-hoc explainability required  
- **Examples:** Random forests, neural networks, NLP classifiers, predictive analytics tools, judicial outcome prediction systems, algorithmic workforce management tools  
- **Tip:** If the paper describes a system using terms like "machine learning", "NLP", "pattern detection", or "predictive analytics" without claiming interpretability, classify as black-box.

---

### 4\. General-purpose systems

Large models pre-trained on massive datasets (typically internet-scale) and adapted to specific tasks via fine-tuning or prompting. These models are not built or managed by the deploying organisation.

- **Key characteristics:** Pre-trained on unknown or large-scale data, adapted via prompting or fine-tuning, general-purpose, opaque training data, generates text/images/other outputs  
- **Examples:** GPT-3, GPT-4, ChatGPT, BERT-based systems, large vision-language models  
- **Tip:** If the paper describes a system as a "large language model", "generative AI", or "GPT", or notes that it generates human-like text from prompts, classify as General-purpose.

---

### 5\. Agentic System

AI systems characterised by autonomy, the ability to interact with their environment, and the capacity to pursue complex goals across multiple steps, often using external tools or APIs.

- **Key characteristics:** Autonomy, tool use, multi-step goal pursuit, environmental interaction, broad generality  
- **Examples:** LLM-based agents with web search or database access, automated workflow systems, multi-agent pipelines  
- **Tip:** If the paper describes a system that acts independently in an environment, uses tools, or completes multi-step tasks without human intervention at each step, classify as Agentic.

---

### 6\. Underspecified

Use this classification when the paper references an AI system without providing sufficient detail to determine its technical nature, or uses AI as a generic concept.

- **Examples:** "AI tools", "automated decision-making systems", "new technologies" without further specification  
- **Tip:** If in doubt between two categories, note both and explain your reasoning. Only use Underspecified when no reasonable classification can be inferred.

**Classification rules:**

* **Be aggressive about "Underspecified"**. If the paper names a concrete system or application (a chatbot, a "decision support tool", a "risk scoring system") but does not provide enough technical detail to determine the system type, classify it as Underspecified even if inference seems plausible. "Sounds like supervised learning" is not sufficient — you need the paper to provide a basis. The bar for moving out of Underspecified is: the paper explicitly names a technique (e.g. "neural network", "logistic regression", "rule-based"), describes a decision logic that makes the type inferrable with high confidence (e.g. "programmed rules to exclude claimants" → Hand-coded), it names a specific tool which is clearly identifiable, or the system is so well-known externally that classification is unambiguous (e.g. facial recognition → Black-box).  
* **"Justifiably generic"** applies in two narrow cases only: (a) the paper is explicitly conceptual about a well-defined class of systems (e.g. "automated decision-making" as a class), or (b) the empirical design itself precludes system-level specificity (e.g. a survey vignette that presents "an AI system" to respondents — you cannot add technical detail to what respondents saw). Even in case (b), the researcher should still demonstrate clarity about what they mean.  
* **For motivation**, use the same typology. "Generic across a class of systems" is appropriate when the paper invokes "AI", "ADM", or "machine learning" without specifying a system — but note this as an imprecision where it matters.

---

### **PUBLIC VALUE DIMENSIONS**

Map all motivations, empirics, and conclusions to one or more of the following dimensions:

### Participation *(Democracy)*

The reference engages with how the AI system affects citizens' or stakeholders' ability to engage with, influence, or be included in public decision-making processes.

- **Sub-values:** Responsiveness, Inclusion, Transparency, Collaboration  
- **Indicators:** Citizens' ability to contest or appeal automated decisions; inclusive design of AI systems; transparency of decision processes to affected parties; collaborative governance of AI deployment  
- **Example:** A paper argues that fully automated benefit decisions eliminate the case-worker interaction through which applicants could raise concerns, reducing meaningful participation.

---

### Procedural justice *(Rule of law)*

The reference engages with the legal and legitimate functioning of governance — whether the AI system operates through fair legal procedures that are suitable, explainable, non-discriminatory, and user-friendly.

- **Sub-values:** Suitability, Explainability, Proportionality, User-friendliness, Disputability, Solution-oriented approach  
- **Indicators:** Whether AI use is legally and procedurally suitable for the decision at stake; whether outputs can be explained to affected individuals; non-discriminatory operation of the system; user-friendliness of AI-mediated services; availability of redress mechanisms  
- **Example:** A paper argues that a black-box risk-scoring tool used in parole decisions cannot provide legally adequate explanations, violating procedural justice requirements.

---

### Human rights *(Rule of law)*

The reference engages with whether the AI system affects individual rights and freedoms, with particular emphasis on data protection and human autonomy.

- **Sub-values:** Non-discrimination, Freedom of expression, Privacy, Human autonomy, Human dignity  
- **Indicators:** Data protection concerns; erosion of individual autonomy through automated nudging or profiling; threats to human dignity in automated interactions; surveillance; chilling effects on expression  
- **Example:** A study finds that an automated content moderation system disproportionately suppresses political speech by minority groups, threatening freedom of expression.

---

### Quality of governance *(Governing capability)*

The reference engages with whether the AI system affects the ability of the governance system to adapt to change and function effectively and competently.

- **Sub-values:** Agility, Expertise, Carefulness, Security, Effectiveness, Efficiency, Independence, Risk awareness  
- **Indicators:** Whether AI enhances or substitutes administrative expertise and competence; agility or adaptability of AI-supported governance; operational security of AI systems; effectiveness and efficiency of AI-supported public services; awareness and management of AI-related risks; independence from vendor lock-in  
- **Example:** A paper examines how adopting a commercial AI platform created problematic dependency on a private vendor, undermining administrative independence.

---

### Responsibility *(Governing capability)*

The reference engages with how the AI system is embedded in the broader system of checks and balances — whether accountability is clear, decisions are verifiable, and ultimate human responsibility is preserved.

- **Sub-values:** Accountability, Verifiability, Human final responsibility, Integrity, Continuity  
- **Indicators:** Clear assignment of responsibility for AI decisions; auditability and verifiability of outputs; ensuring a human remains ultimately responsible for decisions taken with AI support; integrity of AI-supported processes; operational continuity  
- **Example:** A paper argues that when an AI system makes a harmful welfare decision, it is unclear whether responsibility lies with the procuring agency, the vendor, or the individual official, creating a responsibility gap.

---

### None

The reference has no substantive engagement with any of the six PA relevance dimensions.

---

### **SECTION 1: EMPIRICS**

Classify the system(s) the paper actually studies empirically. This section anchors everything else.

For each system or system group:

* Give it a short name  
* Classify it using the taxonomy above, defaulting to Underspecified  
* Write 2–3 sentences explaining your classification, citing what the paper does and does not say  
* Provide 1–3 direct quotes from the paper supporting the classification

If the paper has no empirical component (pure conceptual/literature review), state this explicitly and explain what the closest analogue to an "empirical object" is. A conceptual paper's 'empirical' object is its main motivating example or conceptualisation of AI. A claim about 'AI systems with property P' is classified by which typology layer P uniquely identifies, or as Justifiably Generic if P crosses no affordance thresholds, otherwise Underspecified.

---

### **SECTION 2: MOTIVATION**

Identify **2–5 key motivating strands** — the main reasons the paper gives for why its research question matters. For each strand:

* **One-liner**: A concise description of what the strand is (e.g. "AI tools curbing individual discretion", "opacity of black-box models undermining accountability"). Can be longer if needed. 
* **Public value dimension(s)**: Map to one of the five in the taxonomy above.  
* **Supporting quotes**: 2–5 direct quotes from the paper that anchor this strand.  
* **AI system classification**: Classify the AI system(s) invoked in this motivating strand, using the taxonomy. If the paper is generic here, note "Generic across a class of systems" and flag it.  
* **(A) Mismotivation flag \[0/1\]**: Does this strand motivate with a type of AI system that is meaningfully different from what the paper actually studies (as classified in Section 1)? Flag 1 if yes, with an explanation. The two main failure modes are:  
  * **(a) Imprecision**: the strand invokes AI generically in a way that obscures relevant differences (e.g. treating black-box ML and hand-coded rule systems as the same governance problem).  
  * **(b) Mismotivation proper**: a motivating claim rests on one system type (e.g. Black-box ML) but the paper then studies a different type (e.g. hand-coded rules), making the motivation and the empirics technically mismatched.  
  * **Give leeway** for the common case where a paper invokes "AI" broadly in its framing but the actual gap between what it motivates with and what it studies is minor or conventional. Only flag 1 when the mismatch is consequential — i.e. it would materially change the interpretation of the paper's contribution or its policy implications.

---

### **SECTION 3: CONCLUSIONS / CLAIMS**

Identify the paper's **core conclusions or claims** (typically 3–5). For each:

* **One-liner**: A concise summary of the claim (e.g. "AI tools reduce frontline discretion", "citizen co-design prevents algorithmic harm").  
* **Public value dimension(s)**: Map to the taxonomy above.  
* **Supporting quotes**: 2–4 direct quotes from the paper stating or supporting this conclusion.  
* **(C) Overgeneralisation flag \[0/1\]**: Does the conclusion generalise beyond what the empirics can support, due to technical imprecision? Flag 1 if yes, with an explanation.  
  * The standard here is: does the conclusion apply the finding uniformly to "AI" when the empirics only covered a specific technical subset — and would that matter for the conclusion's validity or policy applicability?  
  * **Give leeway** for the very common case where conclusions mention "AI" generically but the underlying finding is not meaningfully distorted by this. Only flag 1 where the overgeneralisation is consequential: where applying the finding to a different system type (e.g. extending a finding about Black-box ML to hand-coded rule systems, or vice versa) would be a gross mistake, or where the conclusion is the kind that sounds much stronger than the technical evidence warrants.  
  * A conclusion about process or governance (e.g. "citizen involvement improves ethical outcomes") gets somewhat more leeway than a conclusion about technical properties (e.g. "these systems are typically opaque"), since the former may apply across system types more plausibly.

---

### **ADDITIONAL GUIDANCE**

**On technical classification generally:** The typology distinguishes systems by their internal logic and the nature of the inference they perform, not by their application domain. A welfare benefit calculator can be Hand-coded; a welfare risk scorer can be Black-box. A "chatbot" could be Hand-coded (rule-tree), Black-box (LLM/transformer), or Underspecified (no information given). Do not allow domain labels ("AI in healthcare", "policing AI") to substitute for technical classification.

**On the relationship between sections:** The Empirics classification is your anchor. Mismotivation (A) asks whether the motivation is technically consistent with what was studied. Overgeneralisation (C) asks whether the conclusions go beyond what the empirics showed. Both require you to compare against your Empirics classification.

**On quoting:** All quotes must be verbatim from the paper. Do not paraphrase into quote marks.

**On the empirics type field:** Provide a short label only (e.g. "case study", "survey", "vignette experiment", "systematic literature review", "conceptual/framework"). A fuller taxonomy will be applied later.
\end{Verbatim}
\end{document}